\documentclass[runningheads,anonymous]{llncs}
\usepackage{amssymb}
\usepackage{upgreek}
\usepackage{mathtools}
\usepackage{algorithmic}
\usepackage{algorithm}

\usepackage{listings}	
\usepackage{float}
\usepackage{xcolor}
\usepackage[normalem]{ulem}
\usepackage{amssymb}
\usepackage{comment}
\usepackage{tabularx,ragged2e,booktabs}
\usepackage{etoolbox} 
\usepackage{multicol}
\usepackage{multirow}
\usepackage{makecell} 
\usepackage[firstpage]{draftwatermark} 
\usepackage{graphicx} 

\usepackage{cite}  
\newcommand{\SSOE}{\textsc{SSOE}}

\newcommand{\M}{\mathcal{M}}

\newcommand{\Targets}{Targets\newcommand{\Targets}{Targets}

}

\newtheorem{Proof Sketch}{Proof Sketch}

\newboolean{showcomments}
\setboolean{showcomments}{true} 

\ifthenelse{\boolean{showcomments}}{
	\newcommand{\nbc}[3]{
		{\colorbox{#3}{\bfseries\sffamily\scriptsize\textcolor{white}{#1}}}
		{\textcolor{#3}{\sf\small$\langle$\textit{#2}$\rangle$}}}
}{
	\newcommand{\nbc}[3]{}
	
}

\newcommand{\always}{\text{\protect\raisebox{-.5pt}{\ensuremath{\square}}}}
\newcommand{\eventually}{\text{\protect\raisebox{.5pt}{\ensuremath{\lozenge}}}}

\lstdefinelanguage{pseudocode}{
  comment=[l]{//},
  morekeywords=[3]{if,then,else,for,while,do,return,procedure,function,such,that,let,true,false}
}

\begin{document}
\title{
Scaling GR(1) Synthesis via a Compositional Framework for LTL Discrete Event Control\thanks{This version includes correctness modifications with respect to the version published in CAV25}}
\titlerunning{A Framework for Compositional LTL Synthesis of DES}
%

\SetWatermarkAngle{0} 
\SetWatermarkText{\raisebox{20.5cm}{
\hspace{-10cm}
\includegraphics{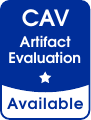}
}}

 \author{Hernan Gagliardi\inst{1} \and
 Victor Braberman\inst{1}\inst{2} \and
 Sebastian Uchitel\inst{1}\inst{2}\inst{3}}
 \authorrunning{H. Gagliardi et al.}
 
 \institute{Universidad de Buenos Aires, Argentina \and
CONICET, Argentina \and
Imperial College London, UK}
\maketitle         

\begin{abstract} 
We present a compositional approach to controller synthesis of discrete event  system controllers with linear temporal logic (LTL) goals. We exploit the modular structure of the plant to be controlled, given as a set of labelled transition systems (LTS), to mitigate state explosion that monolithic approaches to synthesis are prone to. 
Maximally permissive safe controllers are iteratively built for subsets of the plant LTSs by solving weaker control problems. Observational synthesis equivalence is used to reduce the size of the controlled subset of the plant by abstracting away local events. 
The result of synthesis is also compositional, a set of controllers that when run in parallel ensure the LTL goal.
We implement synthesis in the MTSA tool for an expressive subset of LTL, GR(1), and show it computes solutions to that can be up to 1000 times larger than those that the monolithic approach can solve.

\end{abstract}

\section{Introduction}
The problem of automatically constructing control rules from a specification (i.e., synthesis) has been addressed by different areas in Computer Science including  Control of Discrete Event Systems (DES)\cite{ramadge_supervisory_1987}, Reactive Synthesis \cite{pnueli_synthesis} and Automated Planning \cite{nau2004automated}. Each has distinct perspectives on representational and computational aspects. All three use different input languages to provide compact problem descriptions for an underlying semantics that can grow exponentially. These different input languages have led to different algorithmic approaches to achieve more efficient ways to solve synthesis problems.

In this paper, we study control of DES described modularly as the parallel composition of  interacting transition systems. The semantics is a monolithic automaton whose state space can grow exponentially with respect to the number of intervening components. This state explosion makes tractability of DES control problems a  challenge. 
Modularity has enabled compositional reasoning in supervisory control. Notably, in \cite{mohajerani_framework_2014} a compositional framework is described for synthesis of non-blocking supervisory controllers for safety goals. Compositional  DES control for liveness goals (e.g., \cite{Piterman:2006:SRD, DIppolito2008, thistle_control_1995})
has not been studied.

\textit{We present a compositional synthesis framework for modular DES control problems with LTL goals}. 
Our compositional approach does not compose the transition systems describing the plant and then compute a discrete event controller as does~\cite{DIppolito2008}. Instead, roughly, we iteratively pick a subset of plant transition systems, we compute a safe controller for the subset, and replace the subset with a minimized controlled version of the subset that abstracts events not shared with the rest of the plant. Each iteration does a \textit{partial synthesis} of the original control problem.  

The \textit{final result is a controller expressed as a set of controllers} that can be run in parallel to control the original plant, reducing memory footprint and mitigating state explosion of the composed controller.

Our approach is inspired by~\cite{mohajerani_framework_2014} for building modular non-blocking supervisors for finite regular languages.

Moving to $\omega$-regular languages such as those expressed using LTL introduces various challenges. 
First, in contrast to supervisory control for non-blocking supervisors, maximally permissive controllers for LTL control problems are not guaranteed to exist. We compute \textit{safe controllers} for a weaker goal, that allows maximally permissive controllers and supports deferring the construction of a \textit{live controller} that ensures LTL in the final step. Second, although we use the synthesis observational equivalence~\cite{mohajerani_framework_2014} for minimization, the equivalence does not actually preserve LTL equirealizability as it introduces uncontrollable loops which require special considerations. 
Last, in ~\cite{mohajerani_framework_2014} non-blocking conditions are represented modularly with marked states in each plant automaton. Thus, non-blocking can be analysed straightforwardly for subplants. In contrast, LTL is a condition that refers to the behaviour of the composed plant and must be decomposed when analysing subplants. 

We assess if the compositional synthesis can perform better than a monolithic approach in which the plant composition is constructed in full. 
For this we implemented the approach in the MTSA tool~\cite{MTSATool} which is the only tool we are aware of that is available and \textit{natively} supports controller synthesis for modular DES plants. MTSA supports a specific class of LTL formulae referred to as GR(1)\cite{Piterman:2006:SRD}.
\textit{We show that using the same GR(1) explicit state synthesis engine, implemented in MTSA, the compositional approach solves control
problems up to $1000\times$ larger than what a monolithic approach can.} In addition, we show that the compositional framework can sometimes yield more effective results when symbolic synthesis engines such as Strix\cite{meyer_strix_2018} and Spectra\cite{maoz2021spectra} are used for DES controller synthesis via a translation~\cite{majumdar_environmentally-friendly_2019}. 

The remainder of this paper presents background in Section 2, an overview in Section 3, novel definitions and results for compositional synthesis in Section 4, the compositional algorithm in Section 5, and an evaluation in Section 6. We conclude in Sections 7 and 8 with related work and conclusions.

\section{Preliminaries}
We model discrete event systems by composing Labelled Transition Systems.
\begin{definition}[Labelled Transition System]
    A Labelled Transition System (LTS) is a tuple $(S, \Sigma, \xrightarrow[]{}, \hat{s})$ where $S$ is a finite set of states; $\Sigma$ is a finite set of event labels; $\xrightarrow[]{} \, \subseteq S \times \Sigma \times S$ are transitions; and $\hat{s} \in S$ is the initial state.
\end{definition}

\begin{definition}[Parallel composition]
    Let $M_i = (S_i, \Sigma_i, \xrightarrow[]{}_{M_i}, \hat{s}_i)$ with $i\in \{1,2\}$ be two LTSs, their parallel composition is an LTS  $M_1 || M_2 = (S_{1}  \times S_{2}, \Sigma_{1} \cup \Sigma_{2}, \xrightarrow[]{}_{M_1||M_2}, (\hat{s_1}, \hat{s_2}))$, where $\xrightarrow[]{}_{M_1||M_2}$ is the smallest relation that satisfies:
   \begin{itemize}
        \item if $s_1 \xrightarrow[]{l}_{M_1} s_1'$ and $l \in \Sigma_{1} \setminus \Sigma_{2}$, then $(s_1, s_2) \xrightarrow[]{l}_{M_1||M_2} (s_1', s_2)$
        \item if $s_2 \xrightarrow[]{l}_{M_2} s_2'$ and $l \in \Sigma_{2} \setminus \Sigma_{1}$, then $(s_1, s_2) \xrightarrow[]{l}_{M_1||M_2} (s_1, s_2')$
        \item  if $s_1 \xrightarrow[]{l}_{M_1} s_1', s_2 \xrightarrow[]{l}_{M_2} s_2'$ and $l \in \Sigma_{1} \cap \Sigma_{2}$, then $(s_1, s_2) \xrightarrow[]{l}_{M_1||M_2} (s_1', s_2')$
   \end{itemize}
\end{definition}
Parallel composition is associative and commutative. If $\mathcal{M} = \{M_1, M_2, .., M_n\}$ is a set of LTS, we may refer to $M_1 || M_2 || .. ||M_n$ as $\mathcal{M}$ too.

We use $\Sigma^{*}$ for the set of all finite sequences of labels of $\Sigma$, i.e., traces. We say $l \in \Sigma$ is enabled in $s$ if there is $s'$ such that $(s, l, s') \in  \xrightarrow[]{} $. We denote $\xrightarrow[]{}\!(s)$ as the set of enabled events in $s$. An LTS is deadlock-free if all states have outgoing transitions. An LTS is deterministic if there is no state that has two outgoing  transitions on the same event. We use $\xrightarrow[]{}$ to denote the transitive closure of transitions and assume it holds for the empty trace:  $x \xrightarrow[]{\epsilon} x$. We say a (finite or infinite) sequence $w$ of labels in $\Sigma$ is a trace of $M$ if there is an $s'$ such that $\hat{s_M} \xrightarrow[]{w} s'$. 
The concatenation of two traces $s$ and $t$ is written as $st$. For $\Omega \subseteq \Sigma$, we define the natural projection $P_{\Omega}: \Sigma^{*} \xrightarrow[]{} \Omega^{*}$ as the operation that removes from a trace $\pi \in \Sigma^{*}$ all events not in $\Omega$. We use $\Gamma_u, \Gamma_c: S \xrightarrow[]{} S$ to denote the set of states reachable by outgoing uncontrollable (controllable) transitions from an specific state.

\begin{definition}[Controllably reachable]
Given a LTS $M = (S, \Sigma, \xrightarrow[]{}_M, \hat{s})$, $s\in S$ and $l \in \Sigma$, we say that $l$ is \textit{controllably reachable} from $s$ ($s \overset{l}\rightharpoonup_M$) if $(s \xrightarrow[]{l}) \ \vee \ ((\forall s'\in \Gamma_u(s) \ . \ s' \overset{l}\rightharpoonup_{M} ) \ \wedge \ (\Gamma_u(s) \subseteq \emptyset \implies \exists s' \in \Gamma_c(s) \ . \ s' \overset{l}\rightharpoonup_{M}))$. 
\end{definition}

We model controllers with legal LTS that must not block the DES.  

\begin{definition}[Legal LTS]
    Given LTS $M = (S_M, \Sigma, \xrightarrow[]{}_{M}, \hat{s_M})$ and $C = (S_C, \Sigma_C, \xrightarrow[]{}_{C}, \hat{s_C})$ with $\Sigma_u \in \Sigma$. We say that $C$ is a \textit{legal} LTS for $M$ with respect to $\Sigma_u$, if for all $(s_M, s_C)$ such that  $(\hat{s_M}, \hat{s_C}) \xrightarrow[]{}_{M\|C} (s_M, s_C)$ the following holds: $\xrightarrow[]{}_{M||C}\!((s_M, s_C))  \cap \Sigma_u = \, \xrightarrow[]{}_M\!(s_M)\cap \Sigma_u$
\end{definition}

We use Linear Temporal Logic to declaratively specify control requirements. 

\begin{definition}[Linear Temporal Logic (LTL) formula]
    An LTL formula is defined inductively using the standard Boolean connectives and temporal operators $X$ (next), $U$ (strong until) as follows: $\phi := e \ | \ \neg \phi \ | \ \phi \vee \psi \ | \ X \phi \ | \ \phi U \psi$, where $e$ is an atomic event. As usual we introduce $\Diamond$ (eventually), and $\square$ (always) as syntactic sugar. 
    The semantics of an LTL formula is defined via a relation $\models$ between infinite traces and LTL formulae, noted $\pi \models \phi$. 
\end{definition}

We now define discrete event control problems. 

\begin{definition}[Control Problem]
\label{def:control-problem}
Given a set of deterministic LTS $\mathcal{M}$ modelling the plant to be controlled, $\Sigma$ the union of the alphabets of the LTS in $\mathcal{M}$, a set of controllable events $\Sigma_c \subseteq \Sigma$ and an LTL formula $\varphi$ over $\Sigma$ modelling the controller goal, we define $\mathcal{E} = \langle \mathcal{M}, \Sigma_c, \varphi  \rangle$ to be an Control Problem. A deterministic LTS $C$ is a solution for $\mathcal{E}$  (i.e., a controller) if $C$ is \textit{legal} for $\mathcal{M}$ with respect to the set of uncontrolled events $\Sigma_u = \Sigma \setminus \Sigma_c$, $\mathcal{M} || C$ is deadlock-free and for every infinite trace $\pi$ in $\mathcal{M} || C$  we have $\pi \vDash \varphi$. 
\end{definition}

$\mathcal{E}$ is realizable if a solution for $\mathcal{E}$ exists. A solution $C$ is \textit{maximally permissive}~\cite{DBLP:journals/tase/HuangK08} if for any solution $C'$, if $\pi$ is a trace of $C'$ then it is a trace of $C$. 

\begin{definition}[Winning States]
Let $\mathcal{E} = \langle \mathcal{M}, \Sigma_c, \varphi  \rangle$ a  control problem with $\hat{s}$ the initial state of $\mathcal{M}$. We say $s \in S_{\mathcal{M}}$ is a winning state if there is a solution $C$ for $\mathcal{E}$ with initial state $\hat{c}$, a state $c \in S_c$ and a trace $\pi$ such that $(\hat{c},\hat{s}) \xrightarrow[]{\pi}_{C\|M} (c, s)$.

\end{definition}

We focus on a particular form of controller goals referred to as GR(1).

\begin{definition}[Generalised Reactivity(1)~\cite{Piterman:2006:SRD}] A 
GR(1) formula $\varphi$ is an LTL formula restricted with this structure $\varphi = 
\wedge_{i \in n} \ \always\eventually \phi_i \implies \wedge_{j \in m} \ 
\always\eventually \gamma_j$ where $\phi_i$ and $\gamma_j$ are boolean 
combination of events.
\end{definition}

Note that realizable GR(1) control problems may not have a maximally permissive controller. 
An intuition is that a 
controller may delay achieving its goal an unbounded but finite number of times. This  cannot be captured as an LTS.

Also note that that for a realizable GR(1) control problem, if a plant state is losing (i.e., not winning) it will never be traversed when the plant is being controlled. Also note that stripping a plant of its losing states not only preserves realizability but also results in the smallest sub-plant that preserves realizability.

We use an observational equivalence taken from ~\cite{mohajerani_framework_2014}. In~\cite{mohajerani_framework_2014} the relation to reduce the size of the plant, hiding local events $\Upsilon$ while preserving solutions to supervisory control problems with non-blocking goals. The equivalence does not preserve realizability of GR(1) control problems; an adaptation is required.

\begin{definition}[Strong Synthesis observation equivalence (\SSOE)] 
    \label{def:synth-equiv}
    Let $M = (S, \Sigma, \xrightarrow[]{}, \hat{s})$ be an LTS with  $\Sigma = \ \Omega \ \dot{\cup} \ \Upsilon$ (where $\Upsilon$ is the set of labels unique to $M$ relative to the rest of the plant, whereas $\Omega = \Sigma \setminus \Upsilon$), $\Sigma_c$ be the set of controllable events, and $\Sigma_u$ be the set of uncontrollable events. An equivalence relation $\sim_{\Upsilon} \ \subseteq S \times S$ is a \emph{strong synthesis observational equivalence } on $M$ with respect to $\Upsilon$ if the following conditions hold for all $x_1, x_2$ such that $x_1 \sim_{\Upsilon} x_2$:
\vspace{-0.2pt}
    \begin{enumerate}
        \item if $x_1 \xrightarrow[]{u} y_1$ for $u \in \Sigma_u$, then there exist $\pi_1, \pi_2 \in (\Upsilon \cap \Sigma_u)^{*}$ such that $x_2 \xrightarrow[]{\pi_1 P_{\Omega}(u) \pi_2} y_2$, $y_1 \sim_{\Upsilon} y_2$ and $y_2$ is controllably reachable from $x_2$ 
        \item if $x_1 \xrightarrow[]{c} y_1$ for $c \in \Sigma_c$, then there exists a path $x_2 = x_2^{0} \xrightarrow[]{\tau_1} ... \xrightarrow[]{\tau_n} x_2^{n} \xrightarrow[]{P_{\Omega}(c)} y_2$ such that $y_1 \sim_{\Upsilon} y_2$, $y_2$ is controllably reachable from $x_2$  and $\tau_1, .., \tau_n \in \Upsilon$, and if $\tau_{i\in\{1 .. n\}} \in \Sigma_c$ then $x_1 \sim_{\Upsilon} x_2^i$
    \end{enumerate}
\end{definition}

\begin{definition}[Quotient LTS]
Let $M = (S, \Sigma, \xrightarrow[]{}, \hat{s})$ be an LTS, 
and $\sim
\ \subseteq S \times  S$ be an equivalence relation. The \textit{quotient LTS} of $M$
modulo $\sim$
is $M \! \slash \! \!\sim\ =  (S \! \!\ \slash \! \! \sim, \Sigma, \xrightarrow[]{} \! \! \slash \! \! \sim,  [\hat{s}])$ where $S \! \slash \! \! \sim$ is the set of equivalences classes of $\sim$ and $\xrightarrow[]{} \! \! \slash \!  \! \sim =  \{ ([s], \sigma, [s']) \ \cdot \ s \xrightarrow[]{\sigma} s' \}$
\end{definition}

\section{Overview}\label{section:overview}
\label{section:running-example}
In this section we first provide an overview of our approach that does not include an equivalence minimization 
step. We then revisit the 
overview explaining where some aspects of minimization require special treatment.

\vspace{-0.5cm}
\subsubsection{Overview without Minimization.} Consider a control problem $\mathcal{E} = \langle \mathcal{M}, \Sigma_c, \varphi 
\rangle$ where $\mathcal{M}$ is a set of LTS. A monolithic approach to solving the 
problem (e.g., ~\cite{DIppolito2008}) composes all the LTSs in $\mathcal{M}$ and 
then computes a controller. 
Our approach starts with $\mathcal{M}$ and an empty set $\mathcal{C}$ which 
will store \textit{safe controllers} (we discuss them below).  We 
iteratively extract a subset $\mathcal{M}'$ of  $\mathcal{M}$ of size greater than 1, 
compose the LTSs in $\mathcal{M'}$, compute a safe controller for the composition (with 
a weaker property $\varphi'$ and a larger controllable alphabet -- we explain why below). 
and store it in $\mathcal{C}$. In addition, we reintroduce a controlled version of  
$\mathcal{M}'$ to $\mathcal{M}$, 
we refer to this LTS as an \textit{controlled subplant} (we explain these below).  

In each iteration, $\mathcal{C}$ increases by one and $\mathcal{M}$ decreases 
by at least one.
The iteration ends if the subset $\mathcal{M}'$ (which is picked by a 
heuristic) is equal to $\mathcal{M}$, typically when $\mathcal{M}$'s size is 1 or 2. At this point we apply the standard monolithic 
approach for 
$\varphi$ to $\mathcal{M'}$. The controller built in this last step of the algorithm can be 
thought of as a 
\textit{live controller} in contrast to the safe controllers built before.

The \textit{solution} to the original control problem $\mathcal{E}$ is the composition of 
the safe controllers in $\mathcal{C}$ with the live controller computed after the 
iterations.

\paragraph{\textbf{Safe Controllers.}} We provide an intuition of what is referred to as 
safe controllers and why they are computed for a  property $\varphi'$ that is weaker than $\varphi$ and an 
expanded controllable alphabet. We do so by working through an example.

\begin{figure}
    \centering
    \includegraphics[width=.75\linewidth]{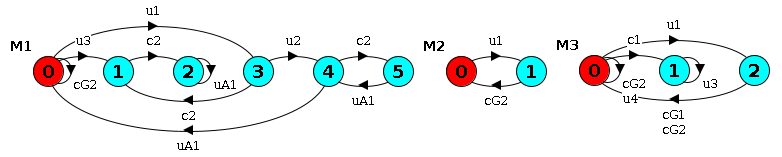}
    \caption{Modular plant for a control problem $\mathcal{E}$ with $ 
    \mathcal{M} = \{M_1, M_2, 
    M_3\}$,  GR(1) goal $\varphi = \always\eventually uA1 \implies
    \always \eventually cG1 \wedge \always \eventually cG2$ and  controllable alphabet 
    $\Sigma_c = \{cG1, cG2, c1, c2\}$.}
    \label{fig:input-machines}
\end{figure}

Consider the control problem $\langle \mathcal{M}, \Sigma_c, \varphi \rangle$  
depicted in Figure~\ref{fig:input-machines}. A monolithic synthesis approach would first 
compose the three LTSs in $\mathcal{M}$  and then compute a controller. The composition (see Fig.10 of the Appendix in the technical report submitted in Arxiv) has 23 states.

\begin{figure}[h!]
\centering
\begin{minipage}[b]{.47\textwidth}
  \centering
  \includegraphics[width=.9\linewidth]{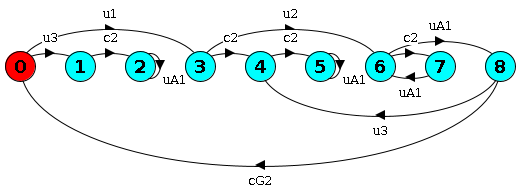}
    \caption{Parallel composition of subplant $\mathcal{M}' = \{M_1, M_2\} \subset 
    \mathcal{M}$, i.e., $M_1 \| M_2$.
    }
    \label{fig:M1_M2}
\end{minipage}\hspace{0.03\textwidth}%
\noindent
\begin{minipage}[b]{.48\textwidth}
    \centering
    \includegraphics[width=.6\linewidth]{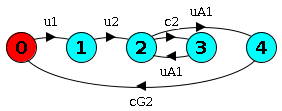}
    \caption{Safe Controller $(\M_1 \| M_2)_{\mbox{\textit{\scriptsize safe}}}$ for 
    subplant $M_1 \| M_2$, 
    $\varphi' =  \always\eventually 
    uA1 
    \implies \top \wedge \always\eventually cG2$, controllable events 
    $\Sigma_c \cup \{u3\}$, and $\mu = \emptyset$.
    }
    \label{fig:safe-controller}
    
\end{minipage}
\end{figure}

Consider a first iteration of our  algorithm in which a heuristic selects the subset 
$\mathcal{M'} = \{M_1, M_2\}$ and computes it  composition (see  Fig. 
\ref{fig:M1_M2}).

We would like to  control $\mathcal{M}'$ to achieve $\varphi$. However, 
this is not possible as $\varphi$ refers to $cG1$ that is  not in the alphabet of the LTS in 
$\mathcal{M}'$. Thus, we weaken $\varphi$  by \textit{projecting} it to the alphabet of 
$\mathcal{M}'$:  $\varphi' = \always\eventually uA1 \implies \top 
\wedge \always\eventually cG2$ .
 
 Now we would like to control $\mathcal{M}'$ to achieve the weaker 
  $\varphi'$. Yet this is not 
  possible because a controller cannot prevent the occurrence of $u3$ in the initial state 
  of $\mathcal{M}'$. 
 But event $u3$ would never occur in the initial state of the plant $\mathcal{M}$ 
 because $M_3 \not\in \mathcal{M'}$ disallows it in its initial state!  Indeed, when 
 computing a controller for $\mathcal{M'}$ we must account for the fact that certain 
 aspects of the problem may be solvable when considering the rest of the plant. For this 
 we \textit{expand the controllable alphabet} to include all uncontrollable events shared 
 with LTS 
 in $\mathcal{M} \setminus \mathcal{M'}$ (i.e., $u3$).

  We now explain why we compute a \textit{safe} controller for the weaker formula $\varphi'$ 
  and the extended set of controllable events:   There may be controllers that can realize 
  $\varphi'$ in different ways (often called strategies). Some of these strategies may be 
  conflicting with restrictions imposed by LTS in $\mathcal{M} \setminus \mathcal{M'}$ 
  or parts of the goal lost in the projection of $\varphi$.  Ideally, we would compute a 
  controller that includes all such strategies, but this is impossible. Liveness goals such as 
  GR(1) do not allow maximally permissive controllers. Instead  of constructing a 
  controller that achieves $\varphi'$ we build one that avoids states in $\mathcal{M}'$ 
  that a controller that guarantees   $\varphi'$ may traverse. In other words, we build a 
  controller that stays in winning states of $\varphi'$. This is a safety property, and if 
  there is a controller, there is a maximally permissive one. 
  
  The safe (maximally permissive) controller for $\mathcal{M'}$, $\varphi$, and 
  $\Sigma_c \cup \{u3\}$, referred to as $(M_1 \| M_2)_{\mbox{\textit{\scriptsize 
  safe}}}$    is 
  shown in 
  Fig.\ref{fig:safe-controller} 
  and is added to $\mathcal{C}$.
  
  \paragraph{\textbf{Controlled Subplant.}} We now provide an intuition of what  
  controlled subplants are. Having computed a safe 
  controller   for $M_1 \| M_2$  for a weaker goal than $\varphi$ we must describe what 
  remains to be controlled. 
  
  A  naive approach would be to add to  $\mathcal{M}$ the controlled version of  
  $M_1 \|    M_2$, in other words $(M_1 \| M_2)_{\mbox{\textit{\scriptsize 
      safe}}} || (M_1 \| M_2)$. However, 
  the $(M_1 \| M_2)_{\mbox{\textit{\scriptsize 
    safe}}}$  was built assuming it controlled $u3$ when in fact it does not. Indeed, the 
    composition would not let the subplant $M_1 \|    M_2$ take a $u3$ transitions   
    from states 0 and 8. 
    
\begin{figure}[h!]
        \centering        \includegraphics[width=.5\linewidth]{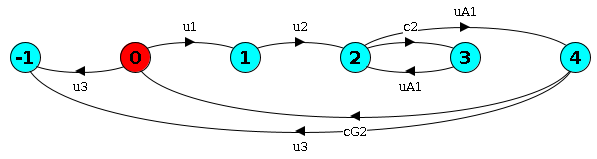}
        \caption{Controlled Subplant $(M_1 || M_2)_{\mbox{\textit{\scriptsize 
          cont}}}$ 
         built from the safe controller $(M_1 || M_2)_{\mbox{\textit{\scriptsize 
          safe}}}$ in
        Fig.~\ref{fig:safe-controller} by adding $u3$ transitions from state 0 and state 4 to a 
        deadlock state}	
        \label{fig:partial-controller}
    \end{figure}

    To adequately model the behavior of the controlled subplant (see 
    Fig.\ref{fig:partial-controller})   
     we must re-introduce  
    $u3$ transitions to the safe controller. Specifically, we must add them to states 0 and 4 
    of the controller, these are the states the controller would be in if it were composed with 
    the subplant and the subplant were in states 0 or 8. Added transitions are sent to a 
    deadlock state because we know that from this point on, it is impossible to achieve 
    $\varphi'$ and thus  impossible to achieve  $\varphi$.  
    
    Indeed, the safe controller with the added transitions and deadlock state represent the 
    safe controller with the assumptions it makes on  when uncontrolled shared  events 
    should not happen. The deadlock state is in a sense abstracting states 1, 2, 4 and 5 of the 
    subplant $M_1 \| M_2$ of Fig.~\ref{fig:M1_M2} as bad states. 
    
\begin{figure}[h!]
              \centering
              \includegraphics[width=.55\linewidth]{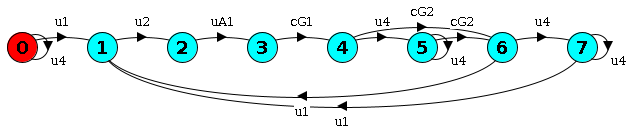}
              \caption{Controller $(M_3 \| (M_1 \| 
                M_2)_{cont})_{live}$ for Fig.11. 
                }
              \label{fig:controllerIt1}
          \end{figure}

  \paragraph{\textbf{After the First Iteration.}}
  At the end of the first iteration of our simplified algorithm we have: $\mathcal{M} = 
  \{M_3, (M_1 \| 
  M_2)_{\mbox{\textit{\scriptsize 
    cont}}}\}$   and $\mathcal{C} =\{ (M_1 \| M_2)_{\mbox{\textit{\scriptsize 
      safe}}}\}$.  As the subplants are required to be of size larger than 1, the only possible   
      $\mathcal{M'} \subseteq \mathcal{M}$ is actually $\mathcal{M}$. Here termination 
      of the iterations kicks in and the algorithm proceeds to build a 
  live controller using a monolithic approach for:  $\langle \{M_3, 
  (M_1 \| M_2)_{\mbox{\textit{\scriptsize 
    cont}}}\}, \Sigma_c, \varphi \rangle$. In other words, it builds the plant $M_3 \| 
    (M_1 
  \| M_2)_{\mbox{\textit{\scriptsize 
    cont}}}$ (see Fig.11 in Appendix of the technical report submitted in Arxiv), 
   computes a controller 
  $(M_3 \| (M_1 \| 
  M_2)_{\mbox{\textit{\scriptsize 
    cont}}})_{\mbox{\textit{\scriptsize 
      live}}}$ (see Fig.~\ref{fig:controllerIt1}) and returns the 
  solution $\mathcal{C} = \{(M_3 \| (M_1 \| 
  M_2)_{\mbox{\textit{\scriptsize 
    cont}}})_{\mbox{\textit{\scriptsize 
      live}}},$ $(M_1 \| M_2)_{\mbox{\textit{\scriptsize 
        safe}}}\}$.

 Note that the largest LTS for which the compositional algorithm built a controller 
 was of size 14, whilst a monolithic approach applied to $M_1 \| M_2 
 \| M_3$ would have 
 had to analyse an LTS of size 22.

\subsubsection{Introducing Minimization.} 
\begin{figure}[h!]
\centering
\begin{minipage}[b]{.45\textwidth}
  \centering
  \includegraphics[width=.55\linewidth]{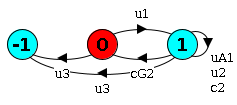}
   \caption{$(M_1 \| 
    M_2)_{\mbox{\textit{\scriptsize 
      cont}}} / \! \!\sim_{\Upsilon}$, a minimized version of 
      $(M_1 \| 
         M_2)_{\mbox{\textit{\scriptsize 
           cont}}}$ with local events $\Upsilon = \{u2, uA1, c2\}$.}
    \label{fig:abstraction}
\end{minipage}\hspace{0.02\textwidth}%
\noindent
\begin{minipage}[b]{.5\textwidth}
    \centering
    \includegraphics[width=1\linewidth]{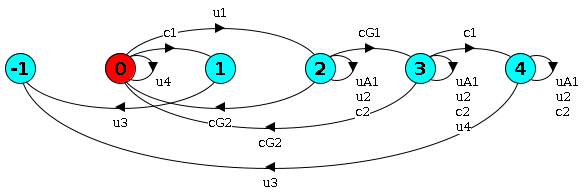}
    \caption{Plant $(M_1 || M_2)^{c} / \sim_{\Upsilon} || M_3$ to be controlled assuming 
    $\square \Diamond \  (\neg  u2  \wedge  \neg uA1 \wedge \neg c2)$)}
    \label{fig:compose_singleton}
\end{minipage}
\end{figure}

The approach to synthesis presented formally in the next sections relies heavily on 
minimization to perform well: It minimizes the controlled subplants before adding them to $\mathcal{M}$.  The minimization procedure preserves the 
synthesis observational equivalence defined in~\cite{mohajerani_algorithm_2012}. We 
apply minimization, as does~\cite{mohajerani_algorithm_2012}, hiding events that are 
local to the controlled subplant (i.e., events that are not part of the alphabet of the remaining LTSs in 
$\mathcal{M}$). In Fig.~\ref{fig:abstraction} we depict the minimized 
version of $(M_1 \| 
M_2)_{\mbox{\textit{\scriptsize 
  cont}}}$ using local events $\Upsilon = \{u2, uA1, c2\}$. Note that minimization may introduce non-determinism that can be solved, as in~\cite{mohajerani_algorithm_2012} using relabeling.  

The approach proceeds by composing $(M_1 \| M_2)_{\mbox{\textit{\scriptsize 
  cont}}} / \!\!\sim_{\Upsilon}$ and 
$M_3$  (see Fig.\ref{fig:compose_singleton}) and 
attempts to build a controller for 
$\varphi$. 
However, such a controller does not exists as the trace $u1.uA1^\omega$ cannot be 
avoided. Yet such a trace is not possible in $(M_1 \| M_2 \| M_3)$ (see 
Fig.10 in Appendix of the technical report submitted in Arxiv).  This mismatch can be  traced back to the minimization procedure that introduces a 
looping transition labelled $uA1$ in state 1 in  $(M_1 \| 
M_2)_{\mbox{\textit{\scriptsize 
  cont}}} /\!\!\sim_{\Upsilon}$ which $(M_1 \| M_2)_{\mbox{\textit{\scriptsize 
  cont}}}$ does not have. Note that these loops are not a problem for compositional control for finite traces such as in~\cite{mohajerani_algorithm_2012}.

    Having introduced uncontrollable looping transitions, they must be kept track of  (we add them to a set $\mu$) and solve control problems in 
    future iterations that assume events in $\mu$  do not occur infinitely (i.e.,  $\always\eventually  \bigwedge_{l\in\mu} \neg l \implies  \varphi$). Although not a GR(1) formula, it can be solved using GR(1) synthesis.

\begin{figure}[h!]
    \centering
    \includegraphics[width=.7\linewidth]{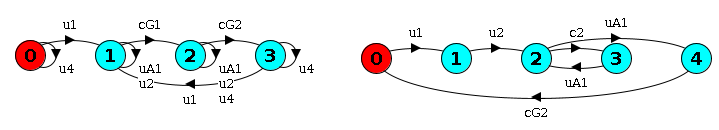}
    \caption{Modular representation of a controller for the discrete event problem 
    $\mathcal{M} = \{M_1, M_2, M_3\}$ to be controlled to satisfy $\varphi = 
    \always\eventually uA1 \implies \always \eventually cG1 \wedge \always 
    \eventually cG2$ with controllable alphabet $\Sigma_c = \{cG1, cG2, c1, c2, c3\}$.
     Controller on the left is a solution to Fig.\ref{fig:compose_singleton}, controller on the 
     right is the safe controller from Fig.~\ref{fig:safe-controller}.} 
    \label{fig:compositional_controller}
\end{figure}

The live controller for $(M_1 \| M_2)_{\mbox{\textit{\scriptsize 
  cont}}} / \sim_{\Upsilon} \| M_3$  and 
property  
$\always\eventually  \bigwedge_{l\in\mu} \neg l \implies \varphi$ is depicted 
on the 
left of Fig. 
\ref{fig:compositional_controller}. This live controller plus the safe 
controller in 
Fig.~\ref{fig:safe-controller} are a modular solution to 
$\mathcal{E}$. 

Note that using minimization, the largest LTS for which synthesis was computed for has 5 
states, while a monolithic approach would have computed a controller for a plant of 22 
states. This exemplifies how state explosion of the composed plant might be avoided using 
our compositional approach.

Subplant minimization can result in a non-deterministic monitor. To remove 
non-determinism we use a the same procedure as in ~\cite{mohajerani_algorithm_2012}.  
Due to space restrictions we do not discuss renaming in this paper, but our implementation and experimentation includes it. We refer the reader to 
~\cite{mohajerani_algorithm_2012} for more information.

\section{Compositional Synthesis}
This section defines a data structure used by the compositional synthesis algorithm, two operations over the data structure, and theorems that enunciate when these operations preserve the semantics of the data structure.

\label{subsection:synthesis-tuple}

During algorithm execution three variables are maintained, ($\mathcal{M}$, $\mathcal{C}$ and $\mu$) that form what we refer to as a 
synthesis tuple. They were discussed in  Section~\ref{section:overview}.

\begin{definition}[Synthesis tuple] \label{def:synthesis-tuple}
    A \textit{synthesis tuple} is tuple  $(\mathcal{M}, \mathcal{C}, \mu)$, where 
    $\mathcal{M}$ is a set of deterministic LTSs modelling a modular plant,  $\mathcal{C}$ is a set of LTSs modelling safe controllers, 
    and $\mu$ is a set of events in the alphabet of $\mathcal{M}$.
\end{definition}

When solving a control problem $\mathcal{E} = \langle \mathcal{M}, \Sigma_c, \varphi 
\rangle$, the algorithm starts with a synthesis tuple of the following form: 
$(\mathcal{M}, \emptyset, \emptyset)$.  The algorithm will iteratively reduce the size of $\mathcal{M}$  while populating 
$\mathcal{C}$ and $\mu$. 
Each intermediate step will yield a synthesis tuple that preserves the control problem to be solved. 
To formalize this notion of preservation we define the set of controllers that are a solution 
to a synthesis tuple. This set of controllers will serve as an invariant. We extend the definition of \textit{composition operator} with respect to the set $\mu$ ($||_{\mu})$ in the appendix (see \ref{def:compOperatorMu}).

\begin{definition}[Controllers of a Synthesis Tuple]
    \label{def:tuple-controller}
    Let $T = (\mathcal{M}, \mathcal{C}, \mu)$ be a synthesis tuple with 
    $\mathcal{C} = \{C_1, \ldots, C_n\}$, 
   $\Sigma_c \subseteq \Sigma$ a controllable alphabet, and $\varphi$ an LTL formula. We define the controllers of a synthesis tuple as
            $Cont_{\Sigma_c, \varphi}(T) = \{ \mathcal{C}_{M} ||_{\mu} C_1 ||_{\mu} \ldots 
            ||_{\mu} C_n \cdot \mathcal{C}_M  \mbox{ is a solution to } \mathcal{E}   \}        $  where $\mathcal{E} = \langle \mathcal{M}, \Sigma_c, 
     \square \Diamond \bigwedge_{l \in 
        \mu} \neg l \!\!\! \implies \!\!\! \varphi 
        \rangle$.

\end{definition}

\begin{remark}
Given the synthesis tuple $T = (\mathcal{M}, \emptyset, \emptyset)$ then the controllers in $Cont_{\Sigma_c, \varphi}(T)$ are the same as those for the control problem  $\mathcal{E} = \langle \mathcal{M}, \Sigma_c, \varphi \rangle$.
\end{remark}

We say that two \textit{synthesis tuples} $T$ and $T'$ \textit{are }\textit{equivalent} with respect to a controllable alphabet $\Sigma_c$ and an LTL formula $\varphi$ if they have the same set of controllers: $T \cong_{\Sigma_c, \varphi} T'$ iff $Cont_{\Sigma_c, \varphi}(T) = Cont_{\Sigma_c, \varphi}(T')$

In the following we introduce two equivalence preserving transformation 
operations over synthesis tuples. 
The compositional 
synthesis algorithm  in Section~\ref{section:algorithm} iteratively 
applies these equivalence preserving transformations. 



\subsection{Composition}
\label{subsection:composition}
This synthesis tuple transformation method  reduces the size of $\mathcal{M}$ in a synthesis tuple by composing in parallel a subset of the LTS in $\mathcal{M}$. It is straightforward to see that parallel composition preserves equivalence of synthesis tuples.

\begin{theorem} [Composition preserves controllers]
    Let $\varphi$ be an LTL formula, $\Sigma_c$ a set of controllable events,  and $T$, $T'$ synthesis tuples $(\mathcal{M}, \mathcal{C}, \mu)$ and $(\mathcal{M'}, \mathcal{C}, \mu)$ respectively. If $\mathcal{M}_1$ and $\mathcal{M}_2$ are a partition of $\mathcal{M}$ such as $\mathcal{M'} = M \cup \mathcal{M}_2$ where $M$ is the parallel composition of the LTS in $\mathcal{M}_1$, then 
    $T \cong_{\Sigma_c, \varphi} 
    T'$
\end{theorem}
The theorem follows from the fact that parallel composition is associative. 

\subsection{Partial Synthesis} 
\label{subsection:partial-synth}

Partial Synthesis is the key transformation operation over synthesis tuples. Given a tuple $(\mathcal{M}, \mathcal{C}, \mu)$ it removes one LTS $M$ from $\mathcal{M}$, computes a safe controller $C$ for a weaker property $\varphi'$ 
and adds the safe controller to $\mathcal{C}$. It also adds a minimized LTS modelling the controlled subplant, i,e. the result of $C$ controlling $M_i$ but hiding local events $\Upsilon$. In addition it updates $\mu$. 

We define the projection of a  formula the alphabet of a subsystem. 



\begin{definition}[Formula projection] \label{formula-projection}
A formula projection is a function $\alpha:LTL\times 2^\Sigma\rightarrow LTL$  such that for any formula $\varphi$ and set $\Sigma' \subseteq \Sigma$ we have $\varphi \implies \alpha(\varphi, \Sigma')$ and $\alpha(\varphi, \Sigma')$ is a formula that only refers to events in $\Sigma'$.
\end{definition}

To achieve completeness, it is essential partial controllers built are maximally permissive. That is, there is a controller that is maximal in terms of traces it allows. 
Maximally permissive controllers in general do not exist. 
Therefore, we construct \textit{{safe controllers}} that avoid reaching a state in the plant from which achieving a the goal is certainly impossible. That is, a controller that keeps the plant in potentially winning states
but does not guarantee traces that satisfy the liveness property.  Liveness is achieved in the algorithm's last step.

\begin{definition}[Safe Controller]
    \label{def:safe-controller}
    Let $\mathcal{E} = \langle \mathcal{M}, \Sigma_c, 
     \varphi \rangle$ be a control problem with $\mathcal{M} = (S, \Sigma, \xrightarrow[]{}, \hat{s})$.  If $\mathcal{E}$ is realizable, we say $(\hat{S}, \Sigma, \xrightarrow[]{}_S, \hat{s})$ is the {safe controller} for $\mathcal{E}$ where
  $\hat{S}$ is a superset of the set of winning states in $S$ for $\mathcal{E}$ and 
$\xrightarrow[]{}_S$ contains all the transitions in $\mathcal{M}$ between states $s \in \hat{S}$. 
\end{definition}

As explained in the overview, a model of the subplant controlled by the safe controller must be reintroduced into $\mathcal{M}$, however this cannot be done by simply introducing their parallel composition as the safe controller will have been built with an extended controllable alphabet. 
We build a \textit{controlled subplant} LTS that models $M$ being controlled by the safe controller $C$ with shared uncontrollable events $\Omega$. Recall from the overview (see Fig. \ref{fig:partial-controller}) that we must add to the safe controller transitions to a deadlock state representing the safe controller's assumptions on when uncontrolled shared events should not happen.  

\begin{definition}[Controlled Subplant] \label{partial-synthesis}
Let $C= (S_C, \Sigma_C, \xrightarrow[]{}_{C},\hat{ {s_C}})$ be a safe controller for  $\mathcal{E} = \langle M, \Sigma_c \cup \Omega, \varphi \rangle$ 
with $M = \langle S_M, \Sigma_M, \xrightarrow[]{}_M, \hat{s_M} \rangle$, $\varphi$ an LTL formula, and $\Omega \subseteq \Sigma_u$ the set of uncontrolled shared events. 
The controlled subplant $M$ by $C$ with respect to $\Omega$ is
     $   \langle S_C \cup \{ \perp \}, \Sigma_C, \xrightarrow[]{}_{}, \hat{s_{C}}  \rangle$
  where $\xrightarrow[]{}_{} = \xrightarrow[]{}_{C} \ \cup \ \{ (s, l, \perp) \ | \ l \in \Omega \wedge s \in S_C \ 
    \wedge \exists s' \in S_M \ \cdot \ s \xrightarrow[]{l}_M s' \ \wedge \forall s' \in S_M \ \cdot (s,l,s') \notin \,
    \xrightarrow[]{}_{C} \}$. 
\end{definition}



We now enunciate how the definitions above can be used to evolve a synthesis tuple by computing a safe controller and adding a minimized controlled subplant. 

Roughly, the theorem states that the solution to a synthesis tuple remains the same if given a tuple $(\mathcal{M}, \mathcal{C}, \mu)$, $(i)$ we take one of the LTS in the plant to be controlled ($M_1 \in \mathcal{M}$), $(ii)$ compute a safe controller for $M_1$, the projection of $\varphi$ to the alphabet of $M_1$, and an controllable alphabet extended with the uncontrollable events shared by $M_1$ and $\mathcal{M}\setminus \{M_1\}$, $(iii)$ add the safe controller to $\mathcal{C}$, $(iv)$ add a minimized version of the subplant $M_1$ controlled by the safe controller to  $\mathcal{M}$ by quotienting it to remove transitions on local events, and $(v)$ add any uncontrollable events on self loops induced by the quotient into $\mu$.  The proof of this theorem is included in the technical report submitted in Arxiv.

\begin{theorem} [Partial synthesis preserves controllers]
\label{PS+SOE_control}
    Given a synthesis tuple $(\mathcal{M}, \mathcal{C}, \mu)$, an LTL formula $\varphi$, and a set $\Sigma_c$ of controllable events.
    Let $M \subseteq \mathcal{M}$ with alphabet $\Sigma_{M} = \Upsilon \dot{\cup} \Omega$ where $\Upsilon$ has the events not shared with any other LTS in $\mathcal{M}\setminus M$ nor in $\varphi$. Let $\varphi'$ be a projection of $\varphi$ to $\Sigma_{M}$. 
    Let $M_{safe}$ be a safe controller for $\langle M, \Sigma_c, \Box\Diamond\bigwedge_{l\in \mu}\neg l \implies \varphi \rangle$.  
    Let $M_{cont}$ be the subplant $M$ controlled by $M_{safe}$ with shared events $\Omega$.

    If $M_{cont} \slash \! \sim_\Upsilon$ is deterministic and $\mu'$ is the set of self-loop events added by $\sim_\Upsilon$ reduction, then  $(\mathcal{M}, \mathcal{C}, \mu) \cong_{\Sigma_c, \varphi} (\mathcal{M} \setminus \{M\} \cup \{M_{cont} \slash \! \sim_\Upsilon\}, \mathcal{C} \cup \{ M_{safe} \}, \mu \cup \mu')$. 
\end{theorem}

Note that the theorem requires  $M_{cont} \slash \! \sim_\Upsilon$ to be deterministic, which is not always the case. Non-determinism can be avoided by renaming labels in $M$ and using distinguishers~\cite{mohajerani_algorithm_2012}. We avoid presenting these due to space restrictions.

\section{Compositional Synthesis Algorithm}\label{section:algorithm}
The algorithm has as inputs a control problem $\mathcal{E} = \langle \mathcal{M}, \varphi, \Sigma_c \rangle$ and a heuristic that returns a subset of the plant of size greater than $1$.

\makeatletter
\renewcommand{\fnum@algorithm}{} 
\makeatother
\noindent

\vspace{-0.75cm}
\noindent
\begin{minipage}[t]{0.48\textwidth} 
\begin{algorithm}[H]
\scriptsize
\caption{\scriptsize{\textbf{CompSynthesis}($\langle \mathcal{M}, \varphi, \Sigma_c \rangle$, heuristic)}}
\begin{algorithmic}[1]
\STATE $\mathcal{C} \gets \{\}$, $\mu \gets \{\}$
\WHILE{$true$}
\STATE sp $\gets$ getSubplant($\mathcal{M}$, heuristic) 
\STATE $\mathcal{M} \gets \mathcal{M} \setminus \text{sp}$
\IF{$|\mathcal{M}|$ == 0}
\STATE $\text{lc} \gets \text{getLiveController}(\langle \text{sp}, \Sigma_c, \varphi \rangle, \mu)$ 
\IF{$\neg$isRealizable(lc)}
\STATE \textbf{return} UNREALIZABLE
\ENDIF
\STATE \textbf{return} $\mathcal{C} \cup \text{\{ lc \}}$
\ENDIF

\STATE PartialSynthesis$(\mathcal{M}, \mathcal{C},  \text{sp}, \Sigma_c, \varphi)$

\ENDWHILE
\end{algorithmic}
\end{algorithm}
\end{minipage}
\hspace{0.0\textwidth} 
\begin{minipage}[t]{0.5\textwidth} 
\begin{algorithm}[H]
\scriptsize
\caption{\scriptsize{$\text{\textbf{PartialSynthesis}}(\mathcal{M}, \mathcal{C}, \text{sp}, \Sigma_c, \varphi)$}}
\begin{algorithmic}[1]

\STATE $\varphi' \gets \text{getProjection}(\varphi, \Sigma_{\text{sp}})$
\STATE $\Sigma_{\varphi'} \gets \text{getAlphabet}(\varphi')$

\STATE $\Upsilon \gets \Sigma_{\text{sp}} \setminus (\Sigma_{\mathcal{M}} \cup \Sigma_{\varphi'})$
\STATE $M_{\mbox{\textit{\scriptsize 
        safe}}} \!\! \gets \text{getSafeController}_{\Upsilon}(\langle \text{sp}, \Sigma_c, \varphi' \rangle, \mu)$
        
\IF{$\neg$isRealizable($M_{\mbox{\textit{\scriptsize 
        safe}}}$)}
\STATE \textbf{return} UNREALIZABLE
\ENDIF
\STATE $M_{\mbox{\textit{\scriptsize 
        cont}}} \!\! \gets \text{getControlledSubplant}_{\Upsilon}(M_{\mbox{\textit{\scriptsize 
        safe}}}, \text{sp})$
\STATE $\mathcal{C} \gets \mathcal{C} \ \cup M_{\mbox{\textit{\scriptsize 
        safe}}}$
\STATE $\langle \text{$M_{\mbox{\textit{\scriptsize 
        cont}}}^{\sim}$}, \mu' \rangle \gets \text{minimize}(\text{$M_{\mbox{\textit{\scriptsize 
        cont}}}$}, \Upsilon)$
\STATE $\mu \gets \mu \cup \mu'$, $\mathcal{M} \gets \mathcal{M} \cup \{\text{$M_{\mbox{\textit{\scriptsize 
        cont}}}^{\sim}$}\}$
\end{algorithmic}
\end{algorithm}
\end{minipage}

\vspace{0.2cm}
The synthesis tuple is initialized and a loop that strictly monotonically reduces the size of $\mathcal{M}$ starts. 
In the loop, the heuristic first selects and removes a subset of $\mathcal{M}$. If $\mathcal{M}$ is not empty, we project $\varphi$ to the alphabet of the subplant, compute the local alphabet ($\Upsilon$)  used to perform minimization, compute the safe controller and the controlled subplant. 
 If there is no solution when computing the safe controller, then the algorithm returns UNREALIZABLE. Otherwise, it adds the safe controller to $\mathcal{C}$ and computes both the quotient of the controlled subplant and the events that the quotient introduces loops for.

The loop terminates when the subplant selected has all LTS in $\mathcal{M}$. At this point we compute a live solution for the remaining control problem and return it with  all previously computed partial controllers.



Procedure \textit{getSubplant} returns a subplant of $\mathcal{M}$ using the \textit{heuristic}. Procedures \textit{getLiveController}
and \textit{getSafeController} solve control problems of the form $\langle sp, \Sigma_c,  \Box\Diamond\bigwedge_{l\in \mu}\neg l \implies \varphi \rangle$ according to Definitions~\ref{def:control-problem} and~\ref{def:safe-controller}. Procedure \textit{getControlledSubplant} returns a controlled subplant according to Definition~\ref{partial-synthesis}, and \textit{minimize} applies minimization based for the equivalence defined in Definition~\ref{def:synth-equiv} and also returns the events for which self-loops have been introduced.

The correctness and completeness argument for the algorithm is as follows. The algorithm encodes the original problem $\mathcal{E}  = \langle \mathcal{M}, \Sigma_c, \varphi \rangle$ as a synthesis tuple $T = (\mathcal{M}, \emptyset, \emptyset)$. The solutions of $T$ by definition are  $Cont_{\Sigma_c, \varphi}(T)$ are exactly those of $\mathcal{E'} = \langle \mathcal{M}, \Sigma_c, \varphi \rangle$ (see Remark 1 after Definition~\ref{def:tuple-controller}). It is straightforward to see that after each iteration of the while loop, the tuple is updated via two solution preserving operations (parallel composition and partial synthesis). The loop terminates because we assume that the heuristic will never infinitely often return subsets of size 1 when $|\mathcal{M}|>1$ size greater than one is replaced in $\mathcal{M}$ with its composition. Thus, the  set of partial controllers  by the algorithm, when composed in parallel is a solution to the original problem.

\section{Evaluation}
The purpose of the evaluation is to assess if a compositional  synthesis of controllers for modular DES plants and LTL goals can perform better than a monolithic approach in which the plant composition is constructed in full. 

For this we implemented compositional synthesis within MTSA~\cite{DIppolito2008}. The only tool we are aware of that is available and \textit{natively} supports controller synthesis for modular DES plants and infinite-trace goals.  By this we mean  means that other tools  do not support describing the system to be controlled as a parallel composition of a set of DES along with some form of specifying  liveness guarantees that are to be achieved. Notably, \cite{supremica} and \cite{eclipse} do modular DES plants, but only allow specifying safety goals. MTSA tool supports GR(1) formulae. We compare MTSA with and without compositional synthesis. 


We also study how reactive synthesis tools perform on translated versions of GR(1) DES control problems, both monolithically and compositionally. Specifically, we study  \textit{Spectra}\cite{maoz2021spectra} (a synthesis tool specialized in GR(1)) and Strix\cite{meyer_strix_2018} (a compositional LTL synthesis tool, winner at SyntComp 2023). Note that we did not implement the compositional approach within \textit{Spectra} and \textit{Strix}, rather we had MTSA call them to measure how much time they take to solve the intermediate control problems compositional synthesis necessitates. 

A replication package including tool and control problems is available in \cite{zenodo}.

\subsection{Compositional Synthesis of DES for  GR(1) goals}

Applying the composition synthesis algorithm 
to control problems where $\varphi$ is a GR(1) formula involves three challenges. 
First and foremost, procedures $getLiveController$ and $getSafeController$ must solve control problems that in principle are not GR(1) (i.e., $\Box\Diamond\bigwedge_{l\in \mu}\neg l \implies \varphi$). However, these problems can actually be solved using a GR(1) synthesis algorithm due to the shape of the subplant under analysis  (the only loops consisting of uncontrollable events that can be found in the subplant are in $\mu$ and are self-loops) and the fact that GR(1) formulae are closed under stuttering.
The procedure is to remove in the plant all self-loop transitions labeled with some $l \in \mu \subseteq \Sigma_u$, compute a controller for $\varphi$, and add to the controller the self-loop transitions previously removed to make it legal. This procedure is correct and complete (see proof in Appendix of the technical report submitted in Arxiv).

\begin{theorem}
    \label{lemma:non-flooding-equivalence}
    Given  control problems $\mathcal{E} =  \langle \mathcal{M'} \setminus \{ (s, l, s) \ | \ l \in \mu \cup \mu' \}, \Sigma_c, \varphi \rangle$ and $\mathcal{E'} =  \langle \mathcal{M'}, \Sigma_c, 
     \square \Diamond (\bigwedge_{l \in (\mu \cup \mu')} \neg l)$ $\implies \varphi \rangle$ where $\mathcal{M'} = \mathcal{M} \setminus \{ M_{sub} \} \cup \{ M_{cont} \slash \! \sim_\Upsilon \}$, $\Sigma_c \subseteq \Sigma$ is a set of controllable events, $\mu \ \text{and} \ \mu'$ are sets of events and $M_{cont}$ is a controlled subplant for $\langle \mathcal{M}_{sub}, \Sigma_c, \square \Diamond (\bigwedge_{l \in \mu} \neg l) \implies \varphi\rangle$ where $\mathcal{
     M}_{sub} \subseteq \mathcal{M}$. If there is a solution $C = (S_c, \Sigma, \xrightarrow[]{}_{c}, \hat{s}_c)$ for control problem $\mathcal{E}$, then $C'$ is a solution for $\mathcal{E'}$ where $C' = (S_{c'}, \Sigma, \xrightarrow[]{}_{c'}, \hat{s}_c)$ is defined as follows:
     \begin{itemize}
         \item $S_{C'} = S_C$
         \item $\xrightarrow[]{}_{C'} =  \xrightarrow[]{}_{C} \cup \ \{ (s, l, s) \ | \ s \in S_c \ \wedge \ l \in \Sigma_u \cap (\mu \cup \mu') \ \wedge \ \exists (s_{\mathcal{M'}}, s_{c}) \in S_{\mathcal{M'}||C} \ \cdot \ (\hat{s_{\mathcal{M'}}}, \hat{s_{C}}) \xrightarrow[]{}_{\mathcal{M'}\|C} (s_{\mathcal{M'}}, s_{C}) \ \wedge \ (s_{\mathcal{M'}}, l, s_{\mathcal{M'}}) \in \xrightarrow[]{}_{\mathcal{M'}} \}$. 
     \end{itemize}
     Also, if there is no solution for $\mathcal{E}$ then there is no solution for $\mathcal{E}'$. 
\end{theorem}

A second challenge is an efficient way to compute a safe controller, that is a controller that keeps a plant in states that are non-losing. Furthermore, that the over-approximation of winning states is sufficiently small to achieve state space reductions that make the compositional approach efficient. This can be achieved for GR(1) formulae straightforwardly as the arena for which the underlying game is solved maps back one-to-one to states of the plant. In addition, the GR(1) synthesis procedure\cite{Piterman:2006:SRD} establishes precisely all winning states of the arena. 

Finally, projecting a GR(1) formula to a weaker formula can be done straightforwardly. In our implementation we use the following procedure: For formulae of the form 
 $\varphi = \wedge_{i \in n} \ \always\eventually \phi_i$ $ \implies \wedge_{j \in m} \ \always\eventually \gamma_j$ where $\phi_i$ and $\gamma_j$ are boolean combination of events in conjunctive normal form and $\Sigma' \in \Sigma$ do the following. For every $l \in \Sigma'$, replace every occurrence of $l$ and $\neg l$ with $F$ in every $\phi_i$ and replace every occurrence of $l$ and $\neg l$ with $\top$ in every $\gamma_j$. It is straightforward to see that this projection yields weaker formulae: $\varphi \implies \alpha(\varphi, \Sigma')$.

\subsection{GR(1) DES Synthesis via Reactive Synthesis} 
To study the performance of GR(1) compositional synthesis for DES via reactive synthesis we implemented a translation of MTSA inputs using an approach based on~\cite{majumdar_supervisory_2021}. 
Nonetheless, we developed two different translations.  The first translates each LTS of the plant and also encodes parallel composition rules. We refer to this translation as \textit{modular}.
The second translates the composed plant, a single LTS. We refer to this translation as \textit{non-modular}. Note that the translation is orthogonal to whether monolithic or compositional synthesis will be applied.

For \textit{Strix} we use a binary-encoding for events and local states to reduce the number of atomic propositions needed. Such an encoding is unnecessary for \textit{Spectra} as it supports variables (and performs binary-encoding internally). 

For both \textit{Strix} and \textit{Spectra} we first evaluate how they perform using the modular and non-modular translations for various instances of the benchmark. We then take the best performing translation for each tool to evaluate how a monolithic synthesis approach compares with a compositional one. 

Monolithic synthesis with \textit{Strix} and \textit{Spectra} is achieved by simply translating the plant of the control problem and executing the reactive synthesis tool. 

To avoid modifying \textit{Strix} and \textit{Spectra} to do compositional synthesis, performance is approximated by running a modified version of the compositional MTSA implementation. Every time MTSA calls its internal GR(1) synthesis procedure with a subplant, we mimick the call by calling the synthesis procedure of \textit{Strix} or \textit{Spectra} with a translation of the subplant. The output of reactive synthesis procedure is discarded, and the result of the native GR(1) synthesis procedure is used. We compute the overall time by subtracting the time spent by internal MTSA synthesis procedure. We refer to these approximations of native compositional implementations as \textit{DES Strix} and \textit{DES Spectra}.

\subsection{Benchmark}
 We use modular DES GR(1) control problems taken from multiple sources. Control problems are parameterizable some scale in two dimensions (number of intervening components, $n$, and number of states per component, $k$) and some scale only in the number of intervening components.

For the case studies scalable in two dimensions, we use all problem families introduced in \cite{Ciolek2020}, modified from supervisory control problems to GR(1) control problems: Air Traffic (AT), Bidding Workflow (BW) and Travel Agency (TA), Transfer Line (TL), and Cat and Mouse (CM). We also used a Robot Search Mission (RS) taken from robotic literature \cite{zudaire_assured_2021}.  
All the n-k instances for these problems are known to be realizable, except for those in AT where $n \leq k$ must hold. Thus, we report separately on instances that are realizable (AT$_R$) and those that are not (AT$_{!R}$). We studied for instances with $n, k \leq 9$.

The one-dimension control problems we use are the classic concurrency Dinning Philosophers (DP) problem and the AI planning problem Gripper (GR)~\cite{DBLP:journals/corr/abs-2305-11014}. Finally, we also use a SYNTECH \cite{maoz2021spectra} case study named Moving Obstacle (MO) originally specified as a reactive synthesis problem (i.e., a turn-based game in which each player can update all its variables in each turn).  We translated MO manually into a modular discrete event problem by developing one LTS for each safety property of the specification plus some additional LTS to replicate the turn-based logic of reactive synthesis control. The maximum value for $n$ varies depending on how large the plant for the control problems grows.

Experiments were run on a computer with an Intel i7-7700 CPU @ 3.60GHz, with 8GB of RAM, and a timeout of 30 min. In addition, we use a simple deterministic heuristic for selecting subplants that always chooses the first two LTS from the vector representation of the set $\mathcal{M}$.


\begin{table*}[t]
    \centering
    {
    \scriptsize
    \begin{tabularx}{\textwidth}{|c|c|c|c|c|c|c|c|c|c|c|c|c|c|c|}
        \cline{1-15}
         & Inst. & \multicolumn{2}{c|}{Solved}  & \multicolumn{10}{c|}{Largest instance solved by both Mono and Comp} & State \\ 
         \cline{5-14}

         & & \multicolumn{2}{c|}{Instances}  &  &   & State &  \multicolumn{2}{c|}{Comp Perf.}  & \multicolumn{2}{c|}{Reduction}  & \multicolumn{3}{c|}{Controller} & space\\

         \cline{3-4} \cline{8-14}
         & & Mono & Comp & n & k & space & States & Time & States & Time & LTSs & States & Red. & lgst. inst.\\
      \cline{1-15}
        
        AT$_R$ & 45 & \textbf{35} & 20 & 4 & 4 & $2^{38}$ & 6244 & 110s & -91\% 
        & -99.63\% 
        & 9 & 6545 & -95.6\%  & $2^{41}$ \\  \cline{1-15}

        AT$_{!R}$ & 36 & \textbf{21} & 15 & 9 & 1 & $2^{42}$ & 17920 & 24s & 33\%  
        & 93\% 
        & -- & -- & --  & $2^{46}$ \\ \cline{1-15}
        
        BW & 81 & 45 & \textbf{50} & 9 & 1 & $2^{22}$ & 1604 & 3s
        & 99.6\% & 96.2\% & 9 & 2499 & 95.6\% & $2^{28}$ \\ 
        \cline{1-15}

        CM & 81 & 20 & \textbf{21} & 4 & 2 & $2^{32}$ & 229993 & 1251s & -51\% & -81.1\% & 7 & 242865 & -91\% & $2^{32}$ \\ 
        \cline{1-15}
        
        RS & 81 & 17 & \textbf{29} & 3 & 3 & $2^{24}$ & 24285 & 230s & 90\% & 84\% & 6 & 212664 & 94.3\% & $2^{30}$ \\ 
        \cline{1-15}

        TA & 81 & 50 & \textbf{71} & 6 & 4 & $2^{50}$ & 6560 & 98s &  97.2\% & 93.4\% & 13 & 7951 & 83.3\% & $2^{67}$ \\ \cline{1-15}
        
        TL & 81 & \textbf{30} & 29 & 3 & 6 & $2^{20}$ & 374993 & 231s & -6\% & -49\% & 6 & 377790 & -99.8\% & $2^{20}$\\ 
        \cline{1-15}

        
        DP & 10 & 4 & \textbf{9} & 5 & - & $2^{20}$ & 185 & 1s & 88\% & 96.8\% & 9 & 391 & 97.3\% & $2^{36}$ \\ \cline{1-15}

        GR & 20 & 8 & \textbf{11} & 8 & - & $2^{29}$ & 378 & 1s & 99.22\% & 96\% & 17 & 853 & 78\% & $2^{38}$ \\ 
        \cline{1-15}
        
        MO & 4 & 3 & \textbf{4} & 6 & - & $2^{100}$ & 5180 & 380s & 59\% & 34\% & 38 & 14011 & 44\% & $2^{127}$ \\ \cline{1-15}

        PC & 10 & 4 & \textbf{10} & 5 & - & $2^{28}$ & 6561 & 220s & 94.8\% & -87\% & 11 & 7152 & 94.3\% & $2^{50}$ \\ \cline{1-15}
        
    \end{tabularx}
    }
    \caption{Experimental Results for Native DES GR(1) Controller Synthesis.}
    \label{table:experimentation}
\label{tab:algorithm_experiment}
\end{table*}
\subsection{Results}

\subsubsection{Monolithic vs. Compositional   GR(1) DES Controller Synthesis.} 

Table \ref{table:experimentation} shows results for all case studies.  The first column has problem family names. The next three aggregate information of entire problem families: we show the total number of instances per problem family (Column Inst) and the number of solved instances by the monolithic (Mono) and compositional (Comp) methods. 

These first three columns show that \textit{for 8 out of 10 control problem families, the compositional method improves over the monolithic one} (we count AT$_R$ and AT$_{!R}$ as the same family). 
In all cases the solved instances one of the methods is a superset of the instances solved by the other. 

All problems that the monolithic approach could not solve were due to running out of memory. Also note that providing more RAM to the monolithic method makes little difference in dealing with the combinatorial growth of the state space resulting from increasing the size of problems using $k$ and $n$. Indeed, doubling memory from 8Gb to 16Gb allows the monolithic approach to add less than 5\% more instances. The two problem families in which monolithic performs better than compositional can be explained by looking at the remaining columns.

From the 5th column onwards we focus on the instance with the largest composed plant that was solved by both monolithic and compositional methods. We first report on the values $n$ and $k$ for the instance and its potential state space (i.e., the product of the sizes of all LTS of the plant).
We then report the performance of the compositional method by showing the number of states of the largest plant which the compositional approach solves a control problem for, and the end-to-end time to compute the final controller. The next two columns report the reduction in the largest control problem to be solved with respect to the monolithic method, and the reduction in time to build the final controller compared to the monolothic approach. Negative values in this column indicate that the monolithic approach did better than the compositional.

\textit{For all but three problem families, when considering the largest instance solved by both methods, the compositional approach was able to significantly reduce the plant on which it computes a live controller.} This is an indication of why for these problem families the compositional method has better performance. The exception being 
AT$_!R$ for which, despite achieving reductions, the total number of instances solved by the compositional approach was not greater. This may indicate some bias of our approach towards realizable problems. 

 The three instances for which we report a negative reduction  (AT$_R$($n=4,k=4$), CM($n=4,k=2$), and TL($n=3, k=6$)) show that state explosion of subplants  can  occur and is a likely cause for poor (or in the case of CM, not significantly better) performance of compositional synthesis.

Notice that treating smaller  (sub)plants does not necessarily mean a reduction in time. Instance PC($n=5$) shows that the compositional method is good at reducing size, but a price is paid in time (mostly spent minimizing subplants).  

The next three columns, show the size of the controller computed by the compositional method. We report in column LTSs how many LTSs is the controller composed of. The States column reports the sum of the states of each of these LTS, other words, a measure of the memory required to store the controller on the computer that will enact the controller. We also show the reduction in memory, measured in states, with respect to the (monolithic) controller computed by the monolithic method. \textit{Note that there are significant reductions in the size of the compositional controller for all case studies except for the three in which there are  intermediate state explosions of subplants.}

The final column shows the theoretical size of the biggest instance solved by either of the methods. Subtracting column 7 (``state space'') we obtain how much bigger was the biggest problem solved by one approach than the biggest of the other: For 4 out of 10 problem families, the compositional approach was able to solve instances at least 3 orders of magnitude  ($\times 2^{10}$) bigger than the monolithic approach. For 3 out 10, the increase in size that the compositional algorithm was able to manage was of at least $2^{6}$ larger. On the other hand for the 3 case studies where monolithic fared better, the largest instances it was able to solve were only $2^{6}$ times bigger.

The final column reports on the state space of the largest instance solved, be it the compositional or the monolithic method. For 4 out of 10 case studies, the compositional approach was able to solve instances at least 3 orders of magnitude  ($\times 2^{10}$) bigger than the monolithic approach. For 3 out 10, the increase in size that the compositional algorithm was able to manage was of at least $2^{6}$ larger. On the other hand for the 3 case studies where monolithic fared better, the largest instances it was able to solve were only $2^{6}$ times  bigger.

The survival plot on the top left of Figure~\ref{fig:cactus-performance} provides a graphical summary of the comparative performance of monolithic vs. compositional GR(1) DES controller synthesis implemented in MTSA. Overall compositional can solve more (and significantly larger) subjects but takes more time than monolithic.

We now discuss possible causes for variable performance of  compositional synthesis across different problem families. We hypothesise that compositional synthesis has opportunities for increased performance when intermediate subsystems can be reduced due to minimization with respect to local events, and when reduction is achieved by reducing reachable states when the subsystem is controlled via partial synthesis.  The main threat to performance is, we believe, having intermediate subsystem state explosion as a result of composing LTS that are very loosely synchronized (worst case being that the composition yields the cartesian product of states). 
Below we analyze two problem families where compositional synthesis performance was worse than monolithic (AT   TL) and one in which it was better (DP). 

For AT, there are two classes of LTS affected by parameters $n$ and $k$: The number of airplane LTS is determined by $n$ and the number of height monitors (that check that at a particular height, there is never more than one airplane)  is determined by $k$. The height monitors interact very little with each other, while each one interacts frequently with all airplanes. Additionally, there is always just one ramp and one response monitor. The response monitor interacts with each airplane on a specific label that is local to that airplane. 

The simple heuristic used composes incrementally height monitors inducing, early on, an intermediate state explosion as their composition yields a state space that is almost the size of the cartesian product of height monitors' states. In addition, there are no local shared events that can lead to intermediate reductions.  
We speculate that a heuristic that starts with the response monitor and adds one airplane at a time will perform better because local shared variables provide opportunities for reduction. Indeed, initial explorations seem to indicate this. 

In instances of TL, there are LTSs representing the machines connected by buffers on a transfer line and an LTS at the end, the Testing Unit (TU), that takes a work piece from the last buffer and accepts it or rejects and returns it to the first buffer for reprocessing.

The naive heuristic we used selects Machines and Buffers in order from the beginning, thus the TU machine is included in the composition towards the end. These intermediate subsystems have permissive controllers because the projected formula (that only talks about events in the TU) is trivial, so they cannot be reduced by partial synthesis. We speculate that a heuristic that starts with TU and adds consecutive machines and buffers from the end towards the beginning of the line will perform better. Initial exploration indicate this is the case. 

In contrast, consider instances of DP, where the compositional approach performs better. The heuristic chooses to add LTSs following the ring architecture defined by the philosophers sitting round the table. Thus, as one more adjacent LTS is added to the subsystem, some new local shared events (between a particular philosopher and a particular fork) can be minimized. Also, since this problem has a formula that refers to events in all philosophers, there are opportunities of minimize by partial synthesis solving control problem for each subsystem.

\vspace{-0.5cm}
\subsubsection{
Monolithic vs Composition GR(1) synthesis via Reactive Synthesis.}

\textit{Strix} performed consistently better when using the non-modular translation than when using the modular one. This provides some evidence that t\textit{he compositional approach Strix has built in (where smaller problems are solved by selecting subsets of formulae) is not able to exploit the modular nature of the DES control problem. } Results for \textit{Spectra} were opposite. The tool performed consistently better when applied to modular translations of the control problems instead of non-modular ones. 
Thus, for the remaining experiments we used modular translations for \textit{Spectra} and non-modular ones for \textit{Strix}.

\begin{figure}[bt]
    \centering
   \centering
  \setlength{\tabcolsep}{2pt} 
  \renewcommand{\arraystretch}{0.9} 
  \begin{tabular}{cc}
    \includegraphics[width=0.4\textwidth]{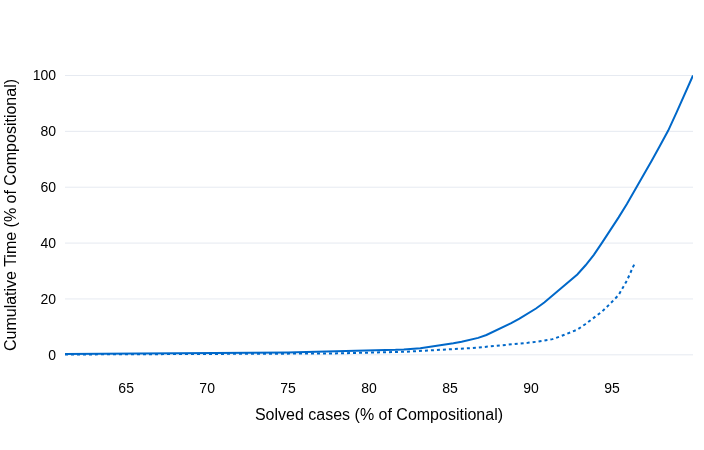} &
    \includegraphics[width=0.4\textwidth]{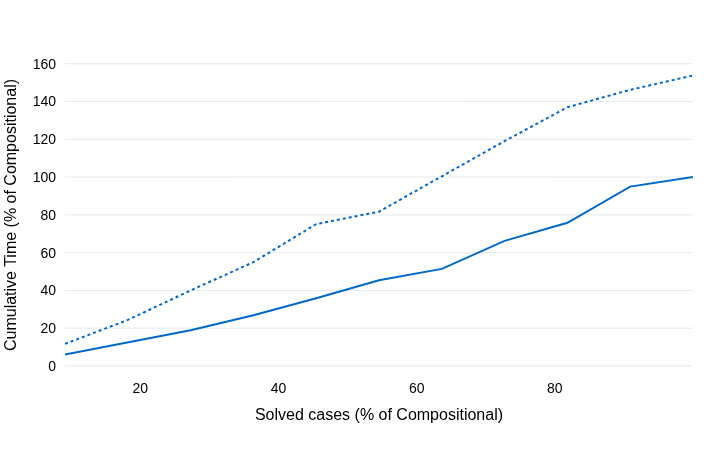} 
    \\
    \includegraphics[width=0.4\textwidth]{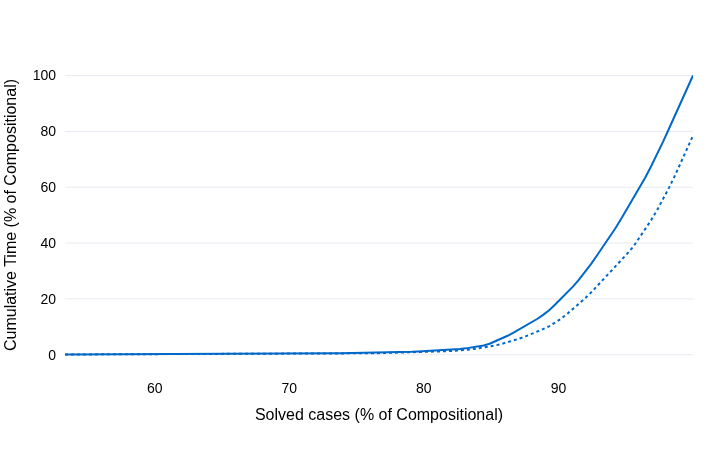} &
    \includegraphics[width=0.4\textwidth]{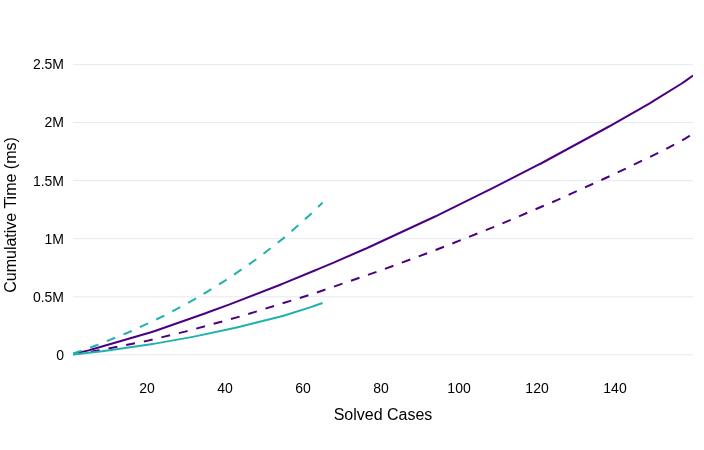}
  \end{tabular}
    \caption{Survival plots for Compositional (solid) and Monolithic (dashed) for \textit{DES MTSA} -top left-, \textit{DES Strix} -top right-,  and \textit{DES Spectra} -bottom left-) relative to compositional performance. Bottom right is for \textit{DES Spectra} divided into problem families for which compositional performs better  (purple) and worse (cyan) than monolithic.}
    \label{fig:cactus-performance}
\end{figure}

 Top right and bottom left survival plots of Fig. \ref{fig:cactus-performance} show comparative performance of compositional vs monolithic \textit{DES Strix} and \textit{DES Spectra}. 
 \textit{DES Strix} solved the same number of cases with both approaches, but the monolithic method took over 50\% more time.
Conversely, bottom left survival plot shows that in \textit{DES Spectra} compositional and monolithic solve the same number of cases, but the monolithic approach is slightly over 20\% faster.

A closer look at \textit{DES Spectra} results reveal that the monolithic tends to perform better on problems with shorter and more centralized formulas (i.e., share alphabet with less LTS).
By splitting problem families based on whether monolithic or compositional \textit{DES Spectra} does better we can see (bottom right of Fig. \ref{fig:cactus-performance}) that for one set of problem families compositional cumulative time grows much less than that of monolithic, while for the other, compositional cumulative time is higher but grows at a closer rate to that of monolithic.




\vspace{-0.5cm}


\subsubsection{Summary.}

The evaluation presented above contributes evidence that compositional synthesis can achieve improvements in scale compared to monolithic DES control, both for native GR(1) DES (explicit state) synthesis, and can achieve improvements in time for DES control via reactive (symbolic state) synthesis. 

For the case of native GR(1) DES synthesis, we observe that performance can degrade if there is intermediate state explosion. Exploring heuristics for choosing subplants that may mitigate this problem. In~\cite{mohajerani_framework_2014} various heuristics are reported to make significant improvements for supervisory control for finite traces. These may also achieve better performance for GR(1) goals.




%

\section{Related Work}
Out of the three classic areas that focus on automatically constructing control rules from a specification,  Discrete Event Control \cite{ramadge_supervisory_1987}, Reactive Synthesis \cite{pnueli_synthesis} and Automated Planning \cite{nau2004automated}, this paper is more related to the first two.

Discrete Event Control was first formulated for *-regular languages \cite{ramadge_supervisory_1987} and then extended to $\omega$-regular languages. In \cite{thistle_control_1995}, Thistle addresses control problems described using a deterministic finite automaton (DFA) equipped with two acceptance conditions for $\omega$-languages, one representing a specification and the other representing an assumption. Supervisors 
are required to generate infinite words that belong to the specification $\omega$-language under the suposition that the environment generates finite traces of the DFA and infinite traces of the 
assumption $\omega$-language. The supervisor is also required to not block the environment from generating some trace in the assumption $\omega$-language.
In \cite{majumdar_supervisory_2021}, this is referred to as a \textit{non-conflicting} requirement. 
In \cite{thistle_control_1995}, Thistle presents a supervisor synthesis algorithm that is polynomial in the number of states of the automaton and exponential in the acceptance condition size.

The control problem we study in this paper can be presented as an instance of the general supervisory control problem for $\omega$-regular languages where the $\omega$-regular language for the  specification is given as an LTL formula. Whereas the $*$-regular language for assumptions is given by a composition of LTS (i.e., the plant) and the $\omega$-regular language for assumptions is the entire set of infinite traces of the plant.



Our work is related to reactive synthesis where the control problem is solely expressed by means of a logical formulae, LTL is commonly used. The system-to-be is a reactive module that shares state variables with the environment. Assumed environment behaviour is also described by means of logical formulae. The underlying interaction model between the controller and the environment differs from that of supervisory control. Reactive synthesis assumes a turn based interaction where each part can change in their turn all the variables it controls. In supervisory control, there are no turns and the environment may win all races.

The relation between supervisory control and reactive synthesis has received much attention in the last years \cite{ehlers_supervisory_nodate}. Notably, \cite{majumdar_supervisory_2021} shows that the non-conflicting/ non-blocking  requirement of supervisory control requires solving games, obliging games, of higher complexity than those classically used for reactive synthesis. The control problem we study does not suffer from this as the non-conflicting requirement is trivially satisfied.

The class of GR(1) LTL formulae have been studied extensively as controllers for them can be computed polynomially~\cite{Piterman:2006:SRD} as opposed to the double exponential complexity of general LTL~\cite{pnueli_synthesis}. GR(1) has been studied extensively as a specification language for reactive systems (e.g.~\cite{menghi_specification_2019, DBLP:journals/tse/MenghiTAPVCG23}) and various variants have been proposed (e.g.,~\cite{majumdar_environmentally-friendly_2019}). 

 GR(1) control of DES as defined in this paper was originally proposed in~\cite{DIppolito2008}, a monolithic synthesis algorithm is defined. 
Later, on-the-fly composition construction has been proposed~\cite{Ciolek2020,DBLP:conf/aips/DelgadoSBU23}.  

The reactive synthesis community has explored compositionality as a way to scale synthesis by identifying and performing independent synthesis tasks~\cite{DBLP:conf/atva/FiliotJR10,DBLP:journals/isse/FinkbeinerGP23,DBLP:conf/atva/FinkbeinerP20}. 
For instance, in~\cite{DBLP:conf/atva/FiliotJR10} LTL synthesis is solved compositionally by first computing winning strategies for each conjunct that appears in the large formula using an incomplete heuristic. The decomposition algorithm of~\cite{DBLP:journals/isse/FinkbeinerGP23} works for LTL and decomposes the formulae into  subspecifications that do not share output variables. A dependency graph used for finding the decomposition-critical propositions that, in turn, are key to soundness and completeness of the approach. 
In~\cite{DBLP:conf/atva/FinkbeinerP20}  decomposition means partitioning system's input and output variables and then applying incremental synthesis of dominant strategies. Dependency graphs must be constructed to identify components and synthesis order that would ensure completeness. On the other hand, our technique works compositionally in terms of environment's safety assumptions (involving input, output and synchronization events) which are described as synchronizing LTS. Goal formula decomposition is a consequence of projection of the formula onto intermediate subsystems. Grouping and ordering of compositions does not affect neither soundness nor completeness (although it may affect performance).


Traditional compositional verification~\cite{DBLP:conf/cav/GrafS90,DBLP:conf/popl/ClarkeGL92} mitigates the state-space explosion by leveraging modular structure of system-under-verification. In that vein, compositional synthesis for plants expressed as finite-state systems has been studied by the community of supervisory control for non-blocking goals~\cite{DBLP:journals/deds/FlordalMFA07,DBLP:journals/deds/MalikMF23}. There, synthesis is  addressed by incrementally composing (and reducing) plant's component. Also, resulting controllers are modular: they are  made up by automata that interact by mean of shared labels.  Our solution is inspired in such works but it addresses  intfinite trace goals for which there is no guarantee for the existence of maximal solutions.

\section{Conclusions and Future Work}
The modular representation of discrete event systems can be exploited by composition synthesis to mitigate the state explosion problem that monolithic synthesis suffers. In this paper we present how compositional synthesis can be applied to LTL. We implement compositional synthesis for GR(1) goals and show that with a very basic heuristic for selecting subplants the compositional synthesis already achieves significant gains with respect to a monolithic method, both in a native DES synthesis tool and using reactive synthesis tools via translations.

Future work must look into avoiding intermediate state explosion by means of smarter heuristics that impact the order in which a plant LTS are composed. An implementation of the compositional framework for LTL requires defining a projection function (which can simply be an existential elimination of  literals) and a method for computing safe controllers. The latter can be trivialized by the conservative decision of not removing any states at all, to ensure that no winning states are lost. However, this is unlikely to provide good performance results. Thus, the challenge with full LTL is to find an effective way of removing a significant portion of subplant losing states.

%
%
%
\section*{Acknowledgements}
This work was partially supported by proyects ANPCYT PICT 2019-1442, 2021-4862, UBACYT 2023-0134, and CONICET PIP 1220220100470CO.

\bibliographystyle{splncs04}
\bibliography{sections/references}

@article{ramadge_supervisory_1987,
	title = {Supervisory control of a class of discrete event systems},
	volume = {25},
	doi = {10.1137/0325013},
	pages = {206--230},
        year={1987},
	journaltitle = {{SIAM} Journal on Control and Optimization},
	shortjournal = {{SIAM} Journal on Control and Optimization},
	author = {Ramadge, Peter and Wonham, Walter},
	date = {1987-01-01},
}

@InProceedings{pnueli_synthesis,
author="Pnueli, Amir
and Rosner, Roni",
editor="Vogt, Frederich H.",
title="A framework for the synthesis of reactive modules",
booktitle="CONCURRENCY 88",
year="1988",
publisher="Springer Berlin Heidelberg",
address="Berlin, Heidelberg",
pages="4--17",
isbn="978-3-540-45999-6"
}

@book{nau2004automated,
  title     = {Automated Planning: Theory and Practice},
  author    = {Ghallab, Malik and Nau, Dana and Traverso, Paolo},
  year      = {2004},
  publisher = {Elsevier},
  address   = {San Francisco, CA},
  isbn      = {978-1-55860-856-6}
}

@article{Ciolek2020,
  author    = {Daniel Alfredo Ciolek and Victor Adrian Braberman and Nicol{\'{a}}s Roque D'Ippolito and Sebastian Uchitel and Sebastian Sardi{\~{n}}a},
  title     = {Compositional Supervisory Control via Reactive Synthesis and Automated Planning},
  journal   = {IEEE Transactions on Automatic Control},
  volume    = {65},
  number    = {8},
  pages     = {3502--3516},
  year      = {2020},
  doi       = {10.1109/TAC.2019.2948270},
  publisher = {Institute of Electrical and Electronics Engineers (IEEE)}
}

@article{mohajerani_framework_2014,
	title = {A {Framework} for {Compositional} {Synthesis} of {Modular} {Nonblocking} {Supervisors}},
	volume = {59},
	issn = {0018-9286, 1558-2523},
	url = {http://ieeexplore.ieee.org/document/6606831/},
	doi = {10.1109/TAC.2013.2283109},
	language = {en},
	number = {1},
	urldate = {2023-09-27},
	journal = {IEEE Transactions on Automatic Control},
	author = {Mohajerani, Sahar and Malik, Robi and Fabian, Martin},
	month = jan,
	year = {2014},
	pages = {150--162},
}

@inproceedings{Piterman:2006:SRD,
 title = {{Synthesis of Reactive(1) Designs}},
 author = {N. Piterman and A. Pnueli and Y. Sa'ar},
 booktitle = {Proc. of the 7th Int. Conf. on Verification, Model Checking and Abstract 
 Interpretation},
 year = {2006},
 volume = {3855},
 publisher = {Springer-Verlag},
}

@misc{menghi_specification_2019,
	title = {Specification {Patterns} for {Robotic} {Missions}},
	url = {http://arxiv.org/abs/1901.02077},
	doi = {10.48550/arXiv.1901.02077},
	urldate = {2023-10-10},
	publisher = {arXiv},
	author = {Menghi, Claudio and Tsigkanos, Christos and Pelliccione, Patrizio and Ghezzi, Carlo and Berger, Thorsten},
	month = jan,
	year = {2019},
	note = {arXiv:1901.02077 [cs]},
}

@article{DBLP:journals/tse/MenghiTAPVCG23,
  author       = {Claudio Menghi and
                  Christos Tsigkanos and
                  Mehrnoosh Askarpour and
                  Patrizio Pelliccione and
                  Gricel V{\'{a}}zquez and
                  Radu Calinescu and
                  Sergio Garc{\'{\i}}a},
  title        = {Mission Specification Patterns for Mobile Robots: Providing Support
                  for Quantitative Properties},
  journal      = {{IEEE} Trans. Software Eng.},
  volume       = {49},
  number       = {4},
  pages        = {2741--2760},
  year         = {2023},
  url          = {https://doi.org/10.1109/TSE.2022.3230059},
  doi          = {10.1109/TSE.2022.3230059},
  bibsource    = {dblp computer science bibliography, https://dblp.org}
}

@article{mohajerani_algorithm_2012,
	title = {An Algorithm for Weak Synthesis Observation Equivalence for Compositional Supervisor Synthesis},
	volume = {45},
	issn = {1474-6670},
	url = {https://www.sciencedirect.com/science/article/pii/S1474667015401776},
	doi = {10.3182/20121003-3-MX-4033.00040},
	series = {11th {IFAC} Workshop on Discrete Event Systems},
	pages = {239--244},
	number = {29},
	journaltitle = {{IFAC} Proceedings Volumes},
	shortjournal = {{IFAC} Proceedings Volumes},
	author = {Mohajerani, Sahar and Malik, Robi and Fabian, Martin},
        year={2012},
	urldate = {2023-12-14},
	date = {2012-01-01},
}

@inproceedings{DIppolito2008,
  author    = {Nicol\'{a}s D’Ippolito and Daniel Fischbein and Marsha Chechik and Sebastian Uchitel},
  title     = {MTSA: The modal transition system analyser},
  booktitle = {Proc. of the Int. Conf. on Automated Software Engineering (ASE)},
  year      = {2008},
  pages     = {},
  publisher = {},
}

@article{zudaire_assured_2021,
	title = {Assured Mission Adaptation of {UAVs}},
	volume = {16},
        year={2021},
	issn = {1556-4665, 1556-4703},
	url = {http://arxiv.org/abs/2107.10173},
	doi = {10.1145/3513091},
	pages = {1--27},
	number = {3},
	journaltitle = {{ACM} Transactions on Autonomous and Adaptive Systems},
	shortjournal = {{ACM} Trans. Auton. Adapt. Syst.},
	author = {Zudaire, Sebastián and Nahabedian, Leandro and Uchitel, Sebastián},
	urldate = {2024-07-31},
	date = {2021-12-31},
	langid = {english},
	eprinttype = {arxiv},
	eprint = {2107.10173 [cs]},
}

@article{maoz2021spectra,
  title={Spectra: a specification language for reactive systems},
  author={Maoz, Shahar and Ringert, Jan Oliver},
  journal={Software and Systems Modeling},
  volume={20},
  pages={1553--1586},
  year={2021},
  publisher={Springer},
  doi={10.1007/s10270-021-00868-z}
}

@article{DBLP:journals/deds/MalikMF23,
  author       = {Robi Malik and
                  Sahar Mohajerani and
                  Martin Fabian},
  title        = {A survey on compositional algorithms for verification and synthesis
                  in supervisory control},
  journal      = {Discret. Event Dyn. Syst.},
  volume       = {33},
  number       = {3},
  pages        = {279--340},
  year         = {2023},
  url          = {https://doi.org/10.1007/s10626-023-00378-8},
  doi          = {10.1007/S10626-023-00378-8},
  bibsource    = {dblp computer science bibliography, https://dblp.org}
}

@inproceedings{DBLP:conf/cav/GrafS90,
  author       = {Susanne Graf and
                  Bernhard Steffen},
  editor       = {Edmund M. Clarke and
                  Robert P. Kurshan},
  title        = {Compositional Minimization of Finite State Systems},
  booktitle    = {Computer Aided Verification, 2nd International Workshop, {CAV} '90,
                  New Brunswick, NJ, USA, June 18-21, 1990, Proceedings},
  series       = {Lecture Notes in Computer Science},
  volume       = {531},
  pages        = {186--196},
  publisher    = {Springer},
  year         = {1990},
  url          = {https://doi.org/10.1007/BFb0023732},
  doi          = {10.1007/BFB0023732},
  bibsource    = {dblp computer science bibliography, https://dblp.org}
}

@inproceedings{DBLP:conf/popl/ClarkeGL92,
  author       = {Edmund M. Clarke and
                  Orna Grumberg and
                  David E. Long},
  editor       = {Ravi Sethi},
  title        = {Model Checking and Abstraction},
  booktitle    = {Conference Record of the Nineteenth Annual {ACM} {SIGPLAN-SIGACT}
                  Symposium on Principles of Programming Languages, Albuquerque, New
                  Mexico, USA, January 19-22, 1992},
  pages        = {342--354},
  publisher    = {{ACM} Press},
  year         = {1992},
  url          = {https://doi.org/10.1145/143165.143235},
  doi          = {10.1145/143165.143235},
  bibsource    = {dblp computer science bibliography, https://dblp.org}
}

@article{DBLP:journals/deds/FlordalMFA07,
  author       = {Hugo Flordal and
                  Robi Malik and
                  Martin Fabian and
                  Knut {\AA}kesson},
  title        = {Compositional Synthesis of Maximally Permissive Supervisors Using
                  Supervision Equivalence},
  journal      = {Discret. Event Dyn. Syst.},
  volume       = {17},
  number       = {4},
  pages        = {475--504},
  year         = {2007},
  url          = {https://doi.org/10.1007/s10626-007-0018-z},
  doi          = {10.1007/S10626-007-0018-Z},
  bibsource    = {dblp computer science bibliography, https://dblp.org}
}

@article{DBLP:journals/isse/FinkbeinerGP23,
  author       = {Bernd Finkbeiner and
                  Gideon Geier and
                  Noemi Passing},
  title        = {Specification decomposition for reactive synthesis},
  journal      = {Innov. Syst. Softw. Eng.},
  volume       = {19},
  number       = {4},
  pages        = {339--357},
  year         = {2023},
  url          = {https://doi.org/10.1007/s11334-022-00462-6},
  doi          = {10.1007/S11334-022-00462-6},
  bibsource    = {dblp computer science bibliography, https://dblp.org}
}

@inproceedings{DBLP:conf/atva/FinkbeinerP20,
  author       = {Bernd Finkbeiner and
                  Noemi Passing},
  editor       = {Dang Van Hung and
                  Oleg Sokolsky},
  title        = {Dependency-Based Compositional Synthesis},
  booktitle    = {Automated Technology for Verification and Analysis - 18th International
                  Symposium, {ATVA} 2020, Hanoi, Vietnam, October 19-23, 2020, Proceedings},
  series       = {Lecture Notes in Computer Science},
  volume       = {12302},
  pages        = {447--463},
  publisher    = {Springer},
  year         = {2020},
  url          = {https://doi.org/10.1007/978-3-030-59152-6\_25},
  doi          = {10.1007/978-3-030-59152-6\_25},
  bibsource    = {dblp computer science bibliography, https://dblp.org}
}

@inproceedings{DBLP:conf/atva/FiliotJR10,
  author       = {Emmanuel Filiot and
                  Naiyong Jin and
                  Jean{-}Fran{\c{c}}ois Raskin},
  editor       = {Ahmed Bouajjani and
                  Wei{-}Ngan Chin},
  title        = {Compositional Algorithms for {LTL} Synthesis},
  booktitle    = {Automated Technology for Verification and Analysis - 8th International
                  Symposium, {ATVA} 2010, Singapore, September 21-24, 2010. Proceedings},
  series       = {Lecture Notes in Computer Science},
  volume       = {6252},
  pages        = {112--127},
  publisher    = {Springer},
  year         = {2010},
  url          = {https://doi.org/10.1007/978-3-642-15643-4\_10},
  doi          = {10.1007/978-3-642-15643-4\_10},
  bibsource    = {dblp computer science bibliography, https://dblp.org}
}

@inproceedings{DBLP:conf/aips/DelgadoSBU23,
  author       = {Tom{\'{a}}s Delgado and
                  Marco S{\'{a}}nchez Sorondo and
                  V{\'{\i}}ctor A. Braberman and
                  Sebasti{\'{a}}n Uchitel},
  editor       = {Sven Koenig and
                  Roni Stern and
                  Mauro Vallati},
  title        = {Exploration Policies for On-the-Fly Controller Synthesis: {A} Reinforcement
                  Learning Approach},
  booktitle    = {Proceedings of the Thirty-Third International Conference on Automated
                  Planning and Scheduling, Prague, Czech Republic, July 8-13, 2023},
  pages        = {569--577},
  publisher    = {{AAAI} Press},
  year         = {2023},
  url          = {https://doi.org/10.1609/icaps.v33i1.27238},
  doi          = {10.1609/ICAPS.V33I1.27238},
  bibsource    = {dblp computer science bibliography, https://dblp.org}
}

@article{DBLP:journals/tase/HuangK08,
  author       = {Jing Huang and
                  Ratnesh Kumar},
  title        = {Directed Control of Discrete Event Systems for Safety and Nonblocking},
  journal      = {{IEEE} Trans Autom. Sci. Eng.},
  volume       = {5},
  number       = {4},
  pages        = {620--629},
  year         = {2008},
  url          = {https://doi.org/10.1109/TASE.2008.923820},
  doi          = {10.1109/TASE.2008.923820},
  bibsource    = {dblp computer science bibliography, https://dblp.org}
}

@article{DBLP:journals/corr/abs-2305-11014,
  author       = {Tom Silver and
                  Soham Dan and
                  Kavitha Srinivas and
                  Joshua B. Tenenbaum and
                  Leslie Pack Kaelbling and
                  Michael Katz},
  title        = {Generalized Planning in {PDDL} Domains with Pretrained Large Language
                  Models},
  journal      = {CoRR},
  volume       = {abs/2305.11014},
  year         = {2023},
  url          = {https://doi.org/10.48550/arXiv.2305.11014},
  doi          = {10.48550/ARXIV.2305.11014},
  eprinttype    = {arXiv},
  eprint       = {2305.11014},
  bibsource    = {dblp computer science bibliography, https://dblp.org}
}

@misc{Zenodo,
  year={2025},
  title = {Replication package: tool and control problems},
  howpublished = {\url{https://zenodo.org/records/13913744}}
}

@article{thistle_control_1995,
	title = {On control of systems modelled as deterministic {Rabin} automata},
	volume = {5},
	copyright = {http://www.springer.com/tdm},
	issn = {0924-6703, 1573-7594},
	url = {http://link.springer.com/10.1007/BF01439153},
	doi = {10.1007/BF01439153},
	language = {en},
	number = {4},
	urldate = {2025-01-09},
	journal = {Discrete Event Dynamic Systems: Theory and Applications},
	author = {Thistle, J. G.},
	month = sep,
	year = {1995},
	pages = {357--381},
}

@misc{majumdar_supervisory_2021,
	title = {Supervisory {Controller} {Synthesis} for {Non}-terminating {Processes} is an {Obliging} {Game}},
	url = {http://arxiv.org/abs/2007.01773},
	doi = {10.48550/arXiv.2007.01773},
	urldate = {2024-12-26},
	publisher = {arXiv},
	author = {Majumdar, Rupak and Schmuck, Anne-Kathrin},
	month = aug,
	year = {2021},
	note = {arXiv:2007.01773 [cs]},
}

@misc{majumdar_environmentally-friendly_2019,
	title = {Environmentally-friendly {GR}(1) {Synthesis}},
	url = {http://arxiv.org/abs/1902.05629},
	doi = {10.48550/arXiv.1902.05629},
	urldate = {2024-12-26},
	publisher = {arXiv},
	author = {Majumdar, Rupak and Piterman, Nir and Schmuck, Anne-Kathrin},
	month = feb,
	year = {2019},
	note = {arXiv:1902.05629 [cs]},
}

@inproceedings{meyer_strix_2018,
	address = {Cham},
	series = {Lecture {Notes} in {Computer} {Science}},
	title = {Strix: {Explicit} {Reactive} {Synthesis} {Strikes} {Back}!},
	isbn = {978-3-319-96145-3},
	shorttitle = {Strix},
	doi = {10.1007/978-3-319-96145-3},
	language = {en},
	booktitle = {Computer {Aided} {Verification}},
	publisher = {Springer International Publishing},
	author = {Meyer, Philipp J. and Sickert, Salomon and Luttenberger, Michael},
	editor = {Chockler, Hana and Weissenbacher, Georg},
	year = {2018},
	pages = {578--586},
}

@article{ehlers_supervisory_nodate,
        year={2017},
	title = {Supervisory {Control} and {Reactive} {Synthesis}: {A} {Comparative} {Introduction}},
	author = {Ehlers and Tripakis},
}

@INPROCEEDINGS{MTSATool,
  author={D'Ippolito, Nicolas and Fischbein, Dario and Chechik, Marsha and Uchitel, Sebastian},
  booktitle={2008 23rd IEEE/ACM International Conference on Automated Software Engineering}, 
  title={MTSA: The Modal Transition System Analyser}, 
  year={2008},
  volume={},
  number={},
  pages={475-476},
  doi={10.1109/ASE.2008.78}}

@article{supremica,
title = {Supremica–An Efficient Tool for Large-Scale Discrete Event Systems},
journal = {IFAC-PapersOnLine},
volume = {50},
number = {1},
pages = {5794-5799},
year = {2017},
note = {20th IFAC World Congress},
issn = {2405-8963},
doi = {https://doi.org/10.1016/j.ifacol.2017.08.427},
url = {https://www.sciencedirect.com/science/article/pii/S2405896317307772},
author = {Robi Malik and Knut Åkesson and Hugo Flordal and Martin Fabian}
}

@InProceedings{eclipse,
author="Fokkink, W. J.
and Goorden, M. A.
and Hendriks, D.
and van Beek, D. A.
and Hofkamp, A. T.
and Reijnen, F. F. H.
and Etman, L. F. P.
and Moormann, L.
and van de Mortel-Fronczak, J. M.
and Reniers, M. A.
and Rooda, J. E.
and van der Sanden, L. J.
and Schiffelers, R. R. H.
and Thuijsman, S. B.
and Verbakel, J. J.
and Vogel, J. A.",
editor="Sankaranarayanan, Sriram
and Sharygina, Natasha",
title="Eclipse ESCET{\texttrademark}: The Eclipse Supervisory Control Engineering Toolkit",
booktitle="Tools and Algorithms for the Construction and Analysis of Systems",
year="2023",
publisher="Springer Nature Switzerland",
address="Cham",
pages="44--52",
isbn="978-3-031-30820-8"
}

\section{Appendix}

\subsection{Composition operator for Solution}\label{def:compOperatorMu} 
Given $C= \{ C_1,..,C_n \}$ a set of LTSs, $L_c$ an LTS where $\Sigma = \Sigma_{C_1} \ \cup \ .. \ \cup \Sigma_{C_n}  \ \cup \ \Sigma_{L_c}$, we define a composition operator $||_{\mu}$ as $(C_1 || \ .. \ || \ C_n \ || \ L_c) \ \setminus \ \text{T}$, where $\text{T} = \{ ((c_1, .., c_n, lc), l, (c_1', .., c_n', lc')) \in \xrightarrow[]{}_{C_1||..||C_n||L_n} \ | \ l \notin f_{C_{i\in 0,..,n}}(c_{i \in 0,..,n}, d) \ \wedge \ d \in \xrightarrow[]{}_{L_N}(lc)  \}$ where $f_M: S \times \Sigma \xrightarrow[]{} \{ \Sigma \}$ is a function defined as $f_M(s, d) = \{ l \ | \ l \in \Sigma_u \ \vee \ (l\pi \ \text{where} \  \pi \ \text{is a controllable path to} \ s' \ \wedge \ s' \xrightarrow[]{d}_{M} ) \}$.

\subsection{Overview}

We show two compositions referred to in the Overview section. The first is the parallel composition of the original modular plant ($M_1\|M_2\|M_3$). The second is the composition of the plant where a subplant has been controlled and minimized ($M_3 \| (M_1 
            \| M_2)_{cont}$). Note the difference in size. 

\begin{figure}[h!]
    \centering
   \includegraphics[width=1\linewidth]{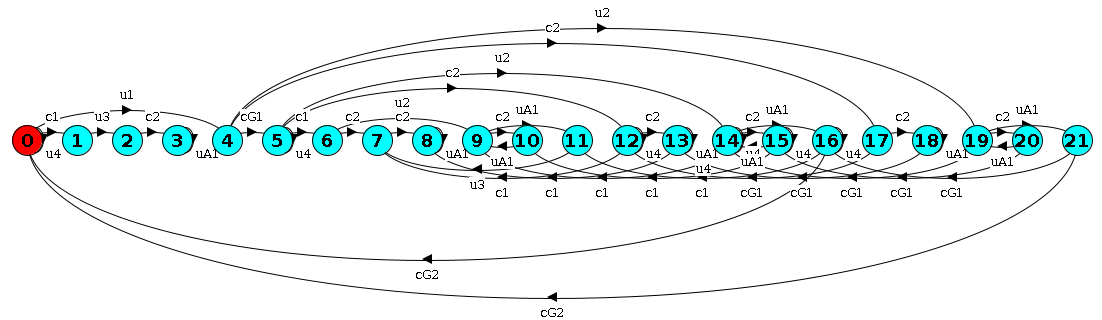}
    \caption{Composed plant  $M_1 \| M_2 \|  M_3$ that a monolithic synthesis approach 
    would 
    have to build to solve the control problem $\mathcal{E}$.} \label{fig:monolithic}
\end{figure}

\begin{figure}[h!]
          \centering
         \includegraphics[width=\linewidth]{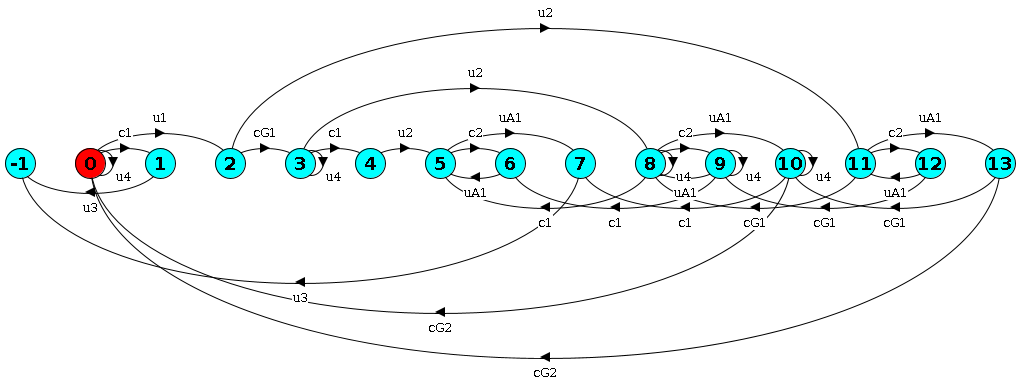}
          \caption{Plant $M_3 \| (M_1 
            \| M_2)_{cont}$. It is nearly half the size of $M_1 \| M_2 
            \| M_3$
            }
          \label{fig:M3||SafeAss}
      \end{figure}

\newpage
\newpage
\subsection{Theorem \ref{lemma:non-flooding-equivalence}}
This Lemma will be used in the proof of Theorem \ref{lemma:non-flooding-equivalence}.
\begin{lemma}
    \label{lemma:selfloop-traces-mu}
    Given a LTS $M$ and $M' = M / \! \sim_\Upsilon$, if there is a trace $\pi \in tr(M')$ and $s\in S_{M'}$ such that $\pi \in \Upsilon^*$ and $s \xrightarrow[]{\pi}_{M'} s$ then $\forall l \in \pi$ we have $s \xrightarrow[]{l}_{M'} s $ 
\end{lemma}
\textbf{Proof.}
We prove by contradiction. Suppose a trace $\pi \in tr(M')$ and $s\in S_{M'}$ such that $\pi \in \Upsilon^*$, $s \xrightarrow[]{\pi}_{M'} s$, and $\exists l \in \pi$ such that $(s, l, s) \notin \xrightarrow[]{}_{M'}$.
Without loss of generality can assume that $\pi = l \pi'$ and $\exists s' \in S_{M'} \cdot s\neq s'$ such that  $s\xrightarrow[]{l}_{M'} \!s'$, $s'\xrightarrow[]{\pi'}_{M'} \!s$. 
Because $s \neq s'$ and $M'$ is minimized respect to $\sim_{\Upsilon}$ we have that $s$ and $s'$ are not equivalent.  However, by Def.~\ref{def:synth-equiv} we have that $s$ and $s'$ are, reaching a contradiction: State $s$ can simulate $s'$ by first going though $l$, and state $s'$ can simulate $s$ by going through $\pi'$.
$\hfill \square$
\\
\\
\noindent The following theorem proves that  a control problem with a formula  of the form $\square \Diamond (\bigwedge_{l \in (\mu \cup \mu')} \neg l) \implies \varphi$ where $\varphi$ is  GR(1)  can be solved as a GR(1) control problem. We use notation $\mathcal{M} \setminus T$ where $\mathcal{M}$ is an LTS and $T$ is a set of transitions over $\mathcal{M}$ to denote the result of removing each transition in $T$ from $\mathcal{M}$. 

\vspace{0.2cm}
\textbf{Theorem \ref{lemma:non-flooding-equivalence}} 
Given a control problem $\mathcal{E} =  \langle \mathcal{M'} \setminus \{ (s, l, s) \ | \ l \in \mu \cup \mu' \}, \Sigma_c, \varphi \rangle$ and $\mathcal{E'} =  \langle \mathcal{M'}, \Sigma_c, 
     \square \Diamond (\bigwedge_{l \in (\mu \cup \mu')} \neg l)$ $\implies \varphi \rangle$ where $\mathcal{M'} = \mathcal{M} \setminus \{ M_{sub} \} \cup \{ M_{cont} \slash \! \sim_\Upsilon \}$, $\Sigma_c \subseteq \Sigma$ is a set of controllable events, $\mu \ \text{and} \ \mu'$ are sets of events and $M_{cont}$ is a controlled subplant for $\langle \mathcal{M}_{sub}, \Sigma_c, \square \Diamond (\bigwedge_{l \in \mu} \neg l) \implies \varphi\rangle$ where $\mathcal{
     M}_{sub} \subseteq \mathcal{M}$. If there is a solution $C = (S_c, \Sigma, \xrightarrow[]{}_{c}, \hat{s}_c)$ for control problem $\mathcal{E}$, then $C'$ is a solution for $\mathcal{E'}$ where $C' = (S_{c'}, \Sigma, \xrightarrow[]{}_{c'}, \hat{s}_c)$ is defined as follows:
     \begin{itemize}
         \item $S_{C'} = S_C$
         \item $\xrightarrow[]{}_{C'} =  \xrightarrow[]{}_{C} \cup \ \{ (s, l, s) \ | \ s \in S_c \ \wedge \ l \in \Sigma_u \cap (\mu \cup \mu') \ \wedge \ \exists (s_{\mathcal{M'}}, s_{c}) \in S_{\mathcal{M'}||C} \ \cdot \ (\hat{s_{\mathcal{M'}}}, \hat{s_{C}}) \xrightarrow[]{}_{\mathcal{M'}\|C} (s_{\mathcal{M'}}, s_{C}) \ \wedge \ (s_{\mathcal{M'}}, l, s_{\mathcal{M'}}) \in \xrightarrow[]{}_{\mathcal{M'}} \}$. 
     \end{itemize}
     Also, if there is no solution for $\mathcal{E}$ then there is no solution for $\mathcal{E}'$.
\\ \\
\textbf{Proof.} We  prove first that if there exists such solution $C$ for $\mathcal{E}$, then $C'$ is a solution for $\mathcal{E'}$:
\begin{itemize}
    \item $C'$ is legal for $\mathcal{M'}$ with respect to $\Sigma_u = \Sigma \setminus \Sigma_c$ \\
    The proof is by contradiction. Suppose that $C'$ is not legal for $\mathcal{M'}$ with respect to $\Sigma_u$. Therefore exists $(s_{C'},s_{\mathcal{M'}})\in S_{C'||\mathcal{M'}}$ and $l \in \Sigma_u$ such that $l \in \xrightarrow[]{}_{\mathcal{M'}}\!(s_{\mathcal{M'}})$ and $l \notin \xrightarrow[]{}_{C'}\!(s_{C'})$. In addition, there is also a trace $\pi \in tr(C') \ \cdot \ \hat{s_{C'}} \xrightarrow[]{\pi} s_{C'}$. Trace $\pi$ may not be in $tr(C)$, however we can build a trace $\pi' \in tr(C)$ that reaches the same state by removing any self loops labeled with $\ l \in \mu \cup \mu'$. This leads us to a contradiction because we assumed that $C$ was legal for $\mathcal{M'} \setminus \{ (s, l, s) \ | \ l \in \mu \cup \mu' \}$ with respect to $\Sigma_u$.
    
    \item $C' || \mathcal{M'}$ is deadlock-free \\ 
     Proven by contradiction. Suppose that $C' || \mathcal{M'}$ is not deadlock-free, so there is a deadlock state $s_d \in S_{C'||\mathcal{M'}}$ and a trace $\pi \in tr(C'||\mathcal{M'})$ such that $(\hat{s_{C'}},\hat{s_{\mathcal{M'}}}) \xrightarrow[]{}_{C'||\mathcal{M'}} s_d$. There are two options:
    \begin{itemize}
        \item $\pi \in tr(C)$ \\
        As we know that $C||\mathcal{M'}$ is deadlock-free, this trivially leads us to a contradiction coming from assumption that $C'||\mathcal{M'}$ is not deadlock free.
        \item $\pi \notin tr(C)$ \\
        If $\pi \notin tr(C)$ there exists a prefix $\pi'l$ of  $\pi$ such that $\pi' \in tr(C) \ \wedge \ \pi'l \notin tr(C)$, and there is a state $\bar{s_C} \in S_C$ such that $\hat{s_C} \xrightarrow[]{}_C \bar{s_C}$. By definition of $\xrightarrow[]{}_{C'}$, $l$ must be a self-loop transition and does not lead to a deadlock state, so this contradicts also initial assumption that $C'||\mathcal{M'}$ is not deadlock-free.
        
        
        
    \end{itemize}

    \item $\forall \pi \in C' || \mathcal{M'},  \pi \vDash \square \Diamond (\bigwedge_{l \in (\mu \cup \mu')} \neg l) \implies \varphi $ \\
    It will be proven by contradiction. Suppose that $\exists \bar{\pi} \in tr( C' || \mathcal{M'}) \ \cdot \ \bar{\pi} \nvDash \square \Diamond (\bigwedge_{l \in (\mu \cup \mu')} \neg l) \implies \varphi$. There are two options:
    
    \begin{itemize}
        \item $\bar{\pi} \in tr(C)$ \\
        As we assume that $\bar{\pi} \nvDash \square \Diamond (\bigwedge_{l \in (\mu \cup \mu')} \neg l) \implies \varphi$, it holds that $\bar{\pi} \nvDash \varphi$. Also we know that $\bar{\pi} \in tr(C)$, so it holds that $\bar{\pi} \vDash \varphi$ because $C$ is solution for $\mathcal{E}$. This result lead us to a contradiction which comes from assuming that exists $\bar{\pi}$.
        \item $\bar{\pi} \notin tr(C)$ \\
        We assume that $\bar{\pi} \nvDash \square \Diamond (\bigwedge_{l \in (\mu \cup \mu')} \neg l) \implies \varphi$, so it holds that $\bar{\pi} \vDash \square \Diamond (\bigwedge_{l \in (\mu \cup \mu')} \neg l) \ \wedge \ \bar{\pi} \nvDash \varphi$. We can build a trace $\bar{\pi}' \subseteq \bar{\pi}$ such that $\bar{\pi}\in tr(C)$ if we remove occurrences of events $l\in \mu\cup\mu'$ of $\bar{\pi}$. As the formula  is stutter-free and there is no formula event present in $\mu\cup\mu'$, we conclude that $\bar{\pi}' \nvDash \varphi$ reaching a contradiction because $\bar{\pi'} \in tr(C)$ and $C$ is a solution for $\mathcal{E}$.
    \end{itemize}
Therefore, it is proven that $\forall \pi \in C' || \mathcal{M'},  \pi \vDash \square \Diamond (\bigwedge_{l \in (\mu \cup \mu')} \neg l) \implies \varphi$.
\end{itemize}
\noindent
What remains is to prove that if there is no solution $C$ for $\mathcal{E}$, then there is none for $\mathcal{E'}$ either. \\
It will be proven by counter reciprocal. I will prove that, if exists a solution $C'$ for $\mathcal{E'}$, then $C$ is a solution for $\mathcal{E}$. 
\begin{itemize}
    \item $C$ is legal for $\mathcal{M'}\setminus \{ (s, l, s) \ | \ l \in \mu \cup \mu' \}$ with respect to $\Sigma_u = \Sigma \setminus \Sigma_c$. \\ 
    The proof is by contradiction. Suppose that $C$ is not legal for $\mathcal{M'} \setminus \{ (s, l, s) \ | \ l \in \mu \cup \mu' \}$ with respect to $\Sigma_u$. Therefore exists $(s_{C},s_{\mathcal{M'}})\in S_{C||\mathcal{M'}}$ and $l \in \Sigma_u$ such that $l \in \xrightarrow[]{}_{\mathcal{M'}}\!(s_{\mathcal{M'}})$ and $l \notin \xrightarrow[]{}_{C}\!(s_{C})$. In addition, there is also a trace $\pi \in tr(C) \ \cdot \ \hat{s_{C}} \xrightarrow[]{\pi}_{C||\mathcal{M'}} s_{C}$. As we know that $tr(C) \subseteq tr(C')$, we conclude that we can reach the same state leading us to a contradiction because we start assuming that $C'$ is legal for $\mathcal{M'}$ with respect to $\Sigma_u$.

    \item $C||\mathcal{M'}$ is deadlock-free\\ 
    Proven by contradiction. Suppose that $C || \mathcal{M'}$ is not deadlock-free, so there is a deadlock state $s_d \in S_{C||\mathcal{M'}}$ and a trace $\pi \in tr(C||\mathcal{M'})$ such that $(\hat{s_{C}},\hat{s_{\mathcal{M'}}}) \xrightarrow[]{}_{C||\mathcal{M'}} s_d$. As we know that $tr(C) \subseteq tr(C')$, then $\pi \in tr(C'||\mathcal{M'})$. On the other hand, we know that $C'||\mathcal{M'} \setminus \{ (s, l, s) \ | \ l \in \mu \cup \mu' \}$ is deadlock-free and this trivially lead us to a contradiction coming from assumption that $C || \mathcal{M'}$ is not deadlock-free.

    \item $\forall \pi \in C || \mathcal{M'} \setminus \{ (s, l, s) \ | \ l \in \mu \cup \mu' \},  \pi \vDash \varphi $ \\
    It will be proven by contradiction. Suppose that $\exists \bar{\pi} \in tr(C||\mathcal{M'} \setminus \{ (s, l, s) \ | \ l \in \mu \cup \mu' \}) \ \cdot \ \bar{\pi} \nvDash \varphi$. As $\bar{\pi} \in tr(\mathcal{M'} \setminus \{ (s, l, s) \ | \ l \in \mu \cup \mu' \})$, it holds trivially that $\bar{\pi} \vDash \square \Diamond (\bigwedge_{l \in (\mu \cup \mu')} \neg l)$. As we assume that $\bar{\pi} \nvDash  \varphi$, then it holds that $\bar{\pi} \nvDash \square \Diamond (\bigwedge_{l \in (\mu \cup \mu')} \neg l) \implies \varphi$.
    On the other hand, we know that $\bar{\pi}\in tr(C')$ (because $tr(C) \subseteq tr(C')$). This lead us to a contradiction because $C'$ is a solution for $\mathcal{E'}$, the contradiction comes from assuming that exists $\bar{\pi}$.
\end{itemize}
$\hfill \square$

\newpage
\subsection{Theorem \ref{lemma:traces-inclusion} }
\noindent Lemmas \ref{lemma:controller-equivalence} and \ref{lemma:traces-inclusion} will be used in the proof of Theorem \ref{PS+SOE_control}.
\begin{lemma}
    \label{lemma:controller-equivalence}
    Given a control problem $\mathcal{E} =  \langle \mathcal{M}, \Sigma_c, \varphi \rangle$, $\mathcal{M} = \{M_1, .., M_n\}$ and $M_{1}^{safe}$ is a Safe Controller for $\mathcal{E'} = \langle M_1, \Sigma_c \cup (\Sigma \setminus \Sigma_{M_1}), \varphi' \rangle$ where $\varphi'$ is a projected formula for $\Sigma_{M_1}$. If $C_M$ is a solution $\mathcal{E}$, then $C_M$ is a solution for $\mathcal{E'}$. 
\end{lemma}
\textbf{Proof.} \\
\begin{itemize}
    \item $C_M$ is legal for $M_1$ with respect to $\Sigma_{u}' = \Sigma \setminus (\Sigma_c \cup (\Sigma \setminus \Sigma_{M_1}))$ \\
    The proof is by contradiction. Suppose that $C_M$ is not legal for $M_1$ with respect to $\Sigma_u' = \Sigma \setminus (\Sigma_c \cup (\Sigma \setminus \Sigma_{M_1}))$. Therefore exists $(s_{C_M}, s_{M_1}) \in S_{C_M||M_1}$ and $l \in \Sigma_u'$ such that $l \in \xrightarrow[]{}_{M_1}(s_{M_1})$ and $l \notin \xrightarrow[]{}_{C_M}(s_{C_M})$. In addition, there is also a trace $\pi \in tr(C_M || M_1)$ such that $(\hat{s_C}, \hat{s_{M_1}}) \xrightarrow[]{\pi}_{C_M||M_1} (s_{C_M},s_{M_1})$. \\
    As we know that $\pi \in tr(C_M)$, we can assume that $\pi \in tr(\mathcal{M})$ and $(\hat{s_C}, \hat{s_{\mathcal{M}}}) \xrightarrow[]{\pi}_{C_M||\mathcal{M}} (s_{C_M}, s_{\mathcal{M}})$ where $l \in \xrightarrow[]{}_{\mathcal{M}}(s_{\mathcal{M}})$ and $l \notin \xrightarrow[]{}_{C_M}(s_{C_M})$. This lead us to a contradiction because $\Sigma_u' \subseteq \Sigma_u$ and $C_M$ is a solution for $\mathcal{E}$, the contradiction comes from assuming that $C_M$ is not legal for $M_1$ with respect to $\Sigma_u' = \Sigma \setminus (\Sigma_c \cup (\Sigma \setminus \Sigma_{M_1}))$. \\
    
    \item $C_M || M_1$ is deadlock-free \\
    The proof is by contradiction. Suppose that $C_M||M_1$ is not deadlock-free, so there is a deadlock state $s_d \in S_{C_M||M_1}$ and a trace $\pi \in tr(C_M||M_1)$ such that $(\hat{s_{C_M}}, \hat{s_{M_1}}) \xrightarrow[]{\pi}_{C_M||M_1} s_d$. As we know that $tr(C_M||M_1) \subseteq tr(C_M||\mathcal{M})$, then $\pi \in tr(C_M||\mathcal{M})$ and this trivially lead us to a contradiction because $C_M$ is a solution for $\mathcal{E}$, the contradiction comes from assuming that $C_M ||M_1$ is not deadlock-free. \\
    
    \item $\forall \pi \in C_M || M_1$, $\pi \vDash \varphi'$ \\
    The proof is by contradiction. Suppose that $\exists \bar{\pi} \in tr(C_M||M_1) \ \cdot \ \bar{\pi} \nvDash \varphi'$. As we know that $tr(C_M||M_1) \subseteq tr(C_M||\mathcal{M})$, then $\pi \in tr(C_M || \mathcal{M})$. This lead us to a contradiction because $\varphi \xrightarrow[]{} \varphi'$, so as $C_M$ is a solution for $\mathcal{E}$, then $C_M$ should satisfy $\varphi'$. The contradiction comes from assuming that $\exists \bar{\pi} \in tr(C_M||M_1) \ \cdot \ \bar{\pi} \nvDash \varphi'$.

\end{itemize}

$\hfill \square$

\begin{lemma}
    \label{lemma:traces-inclusion}
    Given a control problem $\mathcal{E} =  \langle \mathcal{M}, \Sigma_c, \varphi \rangle$, $\mathcal{M} = \{M_1, .., M_n\}$ and $M_{1}^{safe}$ is a Safe Controller for $\mathcal{E'} = \langle M_1, \Sigma_c \cup (\Sigma \setminus \Sigma_{M_1}), \varphi' \rangle$ where $\varphi'$ is a projected formula for $\Sigma_{M_1}$. If $C_M$ is a solution for $\mathcal{E}$, then $P_{\Sigma_{M_{1}}}(tr(C_M)) \subseteq tr(M_{1}^{safe})$. 
\end{lemma}
\textbf{Proof.}
The proof is by contradiction. Suppose that $\exists \pi \in tr(C_M)$ such that $P_{\Sigma_{M_{1}}}(\pi) \notin tr(M_{1}^{safe})$. As $C_M$ is a solution for $\mathcal{E}$, we know that $\pi \vDash \varphi$. Also, as $\varphi \xrightarrow[]{} \varphi'$, it holds that $\pi \vDash \varphi'$. Assuming that $P_{\Sigma_{M_{1}}}(\pi) \notin tr(M_{1}^{safe})$, by definition of safe controller we know that $\exists \ l\in \pi, s \in S_{M_1}$ such that $s \xrightarrow[]{l} s_e$ where $s_e$ is not a winning state (i.e., there is no controller that includes $s_e$). However, by Lemma \ref{lemma:controller-equivalence} we know that $C_M$ is a solution for $\mathcal{E'}$ and $\pi \in tr(C_M)$. This lead us to a contradiction because $P_{\Sigma_{M_{1}}}(tr(C_M))$ goes through winning states of $M_1$, so it must be contained in $tr(M_{1}^{safe})$. The contradiction comes from assuming that $\exists \pi \in tr(C_M)$ such that $P_{\Sigma_{M_{1}}}(\pi) \notin tr(M_{1}^{safe})$.\\

\noindent
The following theorem proves the equirealizability of the most important step that compositional synthesis uses. 

\textbf{Theorem \ref{PS+SOE_control}}  \\
    Given a synthesis tuple $(\mathcal{M}, \mathcal{C}, \mu)$, an LTL formula $\varphi$, and a set $\Sigma_c$ of controllable events.
    
    Let $M \subseteq \mathcal{M}$ with alphabet $\Sigma_{M} = \Upsilon \dot{\cup} \Omega$ where $\Upsilon$ has the events not shared with any other LTS in $\mathcal{M}\setminus M$ nor in $\varphi$. Let $\varphi'$ be a projection of $\varphi$ to $\Sigma_{M}$.

    Let $M_{safe}$ be a safe controller for $\langle M, \Sigma_c, \Box\Diamond\bigwedge_{l\in \mu}\neg l \implies \varphi \rangle$.  
    Let $M_{cont}$ be the subplant $M$ controlled by $M_{safe}$ with shared events $\Omega$.

    If $M_{cont} \slash \! \sim_\Upsilon$ is deterministic and $\mu'$ is the set of self-looping events induced by $\sim_\Upsilon$ reduction, then  $(\mathcal{M}, \mathcal{C}, \mu) \cong_{\Sigma_c, \varphi} (\mathcal{M} \setminus \{M\} \cup \{M_{cont} \slash \! \sim_\Upsilon\}, \mathcal{C} \cup \{ M_{safe} \}, \mu \cup \mu')$.  \\ \\ 
\textbf{Proof.} 
We will use the notation $\mathcal{M'} = \mathcal{M} \setminus \{M\} \cup \{M_{cont} \slash \! \sim_\Upsilon\}$, $T = (\mathcal{M}, \mathcal{C}, \mu)$, $T' = (\mathcal{M} \setminus \{M\} \cup \{M_{cont} \slash \! \sim_\Upsilon\}, \mathcal{C} \cup \{ M_{safe} \}, \mu \cup \mu')$ as syntactic replacement. \\ \\
\textbf{$\xrightarrow{}$)} Given $S \in Cont_{\Sigma_c, \varphi}(T)$, we prove that $S \in Cont_{\Sigma_c, \varphi}(T')$. \\
This will be proven by existence of $\hat{C_{M}}$ such that is a solution for $\mathcal{E'} = \langle \mathcal{M'}, \Sigma_c, \square \Diamond \bigwedge_{l \in \mu \cup \mu'} \neg l \implies \varphi \rangle$ and $\hat{C_{M}} || M_{safe} || C = S$. \\
Since $S = C || C_M$ by definition where $C_M$ is a solution for $\mathcal{E} = \langle M, \Sigma_c, \square \Diamond \bigwedge_{l \in \mu} \neg l \implies \varphi \rangle$. We  demonstrate $\hat{C_M}$ exists using $\hat{C_M} = C_M$ and proving that:
\begin{enumerate}
    \item $C_M$ is a solution for $\mathcal{E'}$.
    \begin{enumerate}
        \item \textit{(formula satisfiability)} We  prove that 
        $\forall \pi \in (C_M || \mathcal{M'}) \ . \ \pi \vDash \square \Diamond \bigwedge_{l \in \mu \cup \mu'} \neg l \implies \varphi$ by contradiction. Suppose $C_M$ is not a solution for $\mathcal{E'}$ and there is a trace $\pi \in tr(C_M || \mathcal{M'})$ such that $\pi \nvDash 
        \square \Diamond \bigwedge_{l \in \mu \cup \mu'} \neg l \implies \varphi$. Also, it holds that $\pi \in C_M$.\\
        Since $C_M$ is a solution for $\mathcal{E}$, then $\forall \pi \in tr(C_M || \mathcal{M}) \ . \ \pi \vDash \square \Diamond \bigwedge_{l \in \mu} \neg l \implies \varphi$. \\
        Note that any trace $\pi$ that $\pi \vDash \square \Diamond \bigwedge_{l \in \mu} \neg l \implies \varphi$, it holds that \\ $\pi \vDash \square \Diamond \bigwedge_{l \in \mu \cup \mu'} \neg l  \implies \varphi$ because you are strengthening the assumption. Then, $\forall \pi \in tr(C_M) \ . \ \pi \vDash \square \Diamond \bigwedge_{l \in \mu \cup \mu'} \neg l \implies \varphi$ which contradicts the original assumption. \\
        Then, $\forall \pi \in (C_M || \mathcal{M'}) \ . \ \pi \vDash \square \Diamond \bigwedge_{l \in \mu \cup \mu'} \neg l \implies \varphi$.
        
        \item $C_M$ is legal for $\mathcal{M'}$ with respect to $\Sigma_u \cup (\mu \cup \mu')$.

         Proven by contradiction. \\
        Suppose $C_M$ is not legal for $\mathcal{M'}$ with respect to $\Sigma_u \cup (\mu \cup \mu')$. Then, $\exists \pi \in tr(C_M)$ such that $\exists l \in (\Sigma_u \cup (\mu \cup \mu')) \wedge \pi l \in tr(\mathcal{M'}) \wedge \pi l \notin tr(C_M)$. \\
        I separate into two cases: $l \in \Sigma_u \cup \mu$ or $l \in \mu'$
        \begin{enumerate}
            \item $l \in \Sigma_u \cup \mu$ \\
            This is not possible since $C_M$ is legal for $\mathcal{M}$ with respect to $\Sigma_u \cup \mu$ and $tr(\mathcal{M}) \subseteq tr(\mathcal{M'})$.
            \item $l \in \mu'$ \\
            By Lemma \ref{lemma:traces-inclusion}, if $\pi l \notin P_{M_{safe}}(tr(C_M))$, then $\pi l \notin tr(M_{safe})$. \\
            Then, by construction, $\pi \notin tr(M_1).$ \\
            Then, because $tr(\mathcal{M}) \subseteq tr(\mathcal{M'})$, $\pi l \notin tr(\mathcal{M'}).$ \\
            Which enter in contradiction with original assumption. Then, $C_M$ is legal for $\mathcal{M'}$ with respect to $\Sigma_u \cup (\mu \cup \mu')$.
        \end{enumerate}

        

        \item ($C_M || \mathcal{M'})$ is deadlock-free. \\
        Proven by contradiction. \\ 
        We suppose that $C_M || \mathcal{M'}$ has a deadlock state $(c, \mathcal{M'})$ reachable from initial state by a trace $\pi$. 
        Since $C_M || \mathcal{M}$ is deadlock free, there exists a label $l \in \Sigma_\mathcal{M}$ which is enabled in $(c, m)$, where $(c,m) \in S_\mathcal{M}$ is an state reachable by trace $\pi$.  
        Since $tr(\mathcal{M}) \subseteq tr(\mathcal{M'})$, it is not possible that $(c, m)$ has an enabled event which is disabled in $(c, \mathcal{M'})$. \\
        Then, ($C_M || \mathcal{M'})$ is deadlock-free.
        
    \end{enumerate}
    Then, $C_M$ is a valid solution for $\mathcal{E'}$. \\
    
    \item $C_M || M_{safe} || C = S$ \\
    This is correct iff $C_M || M_{cont} = C_M$ and it is correct because $tr(C_M |_{\Sigma_{M_{cont}}}) \subseteq tr(M_{cont})$, which is true by Lemma \ref{lemma:traces-inclusion}.
\end{enumerate}

\textbf{$\xleftarrow[]{}$)} Given $S \in Cont_{\Sigma_u, \varphi}(T')$, We prove that $S \in Cont_{\Sigma_u, \varphi}(T)$  by existence of $C_{M}$ such that is a solution for $\mathcal{E} = \langle \mathcal{M}, \Sigma_c, \square \Diamond \bigwedge_{l \in \mu} \neg l \implies \varphi \rangle$ and $C_M || C = S$. Since $S = \hat{C_M} || C || M_{safe}$ where $\hat{C_M}$ is a solution for  $\mathcal{E'} = \langle \mathcal{M'}, \Sigma_c, \square \Diamond \bigwedge_{l \in \mu \cup \mu'} \neg l \implies \varphi \rangle$.
We demonstrate $C_M$ exists using $C_M = \hat{C_{M}} || M_{safe}$ and prove that:

\begin{enumerate}
    \item $\hat{C_{M}} || M_{safe}$ is a solution for $\mathcal{E} = \langle \mathcal{M}, \Sigma_c, \square \Diamond \bigwedge_{l \in \mu} \neg l \implies \varphi \rangle$.

    \begin{enumerate}
        \item (\textit{formula satisfiability}) We will prove that $\forall \pi \in tr(\hat{C_{M}} || M_{safe}||\mathcal{M}) \ \cdot \ \pi \vDash \square \Diamond \bigwedge_{l \in \mu} \neg l \implies \varphi$. Since $\hat{C_M}$ is a solution for $\mathcal{E'}$, then $\pi \vDash \square \Diamond \bigwedge_{l \in \mu \cup \mu'} \neg l \implies \varphi$. We must guarantee that $\pi \vDash \square \Diamond \bigwedge_{l \in \mu'} \neg l$. As $\mu'$ is a set of self-loop labels induced by minimization, we know that there are not present as self-loop in $\mathcal{M}||M_{safe}$ because there are prior to last minimization. Therefore we can conclude that $\pi \vDash \square \Diamond \bigwedge_{l \in \mu'} \neg l$.
        Then, $\pi \vDash (\square \Diamond \bigwedge_{l \in \mu \cup \mu'} \neg l \implies \varphi) \wedge \pi \vDash (\square \Diamond \bigwedge_{l \in \mu'} \neg l)$ implies $\pi \vDash \square \Diamond \bigwedge_{l \in \mu} \neg l \implies \varphi$. \\
        Finally, we can say that $\forall \pi \in tr(\hat{C_M} || M_{cont} || \mathcal{M}) \ . \ \pi \vDash \square \Diamond \bigwedge_{l \in \mu} \neg l \implies \varphi$. \\
       
        \item $\hat{C_{M}} || M_{safe}$ is legal for $\mathcal{M}$ with respect to $\Sigma_u \cup \mu$. \\
        We prove this by contradiction. Then, we suppose that exists an state $(\hat{c}, m_{safe}, m) \in S_{\hat{C_{M}} || M_{safe}||\mathcal{M}}$ and an event $l \in \Sigma_u \cup \mu$ that is enabled for $m$ in over $\mathcal{M}$ transitions but not in $\hat{c}$ or $m_{safe}$. \\
        We know there is a trace $\pi \in tr(\hat{C_M} || M_{safe})$ that reaches mentioned state. \\
        Then, $\pi \in tr(\hat{C_M}) \wedge \pi \in tr(M_{safe})$. \\
        Also, as $tr(\mathcal{M}) \subseteq tr(\mathcal{M'})$, $\pi \in tr(\mathcal{\mathcal{M'}})$ \\
        There are two options that could explain state $(\hat{c}, m_{safe}, m)$ blocks a $l$ event.
        \begin{itemize}
            \item We suppose that $\hat{c}$ has not enabled $l$ event, but it is impossible because $\hat{C}$ is legal for $\mathcal{M'}$ with respect to $\Sigma_u \cup \mu \cup \mu'$.
            \item We suppose that $m_{safe}$ has not enabled $l$ event. Also, we know $\pi \in tr(M)$ and reaches an state where $l$ is enabled in $M$. There is a same reasoning for $M_{safe}$ because it is (by construction) safe and it does not block any uncontrollable event neither an event which is local for respect another component in $\mathcal{M}$.
        \end{itemize}
        Then, $\hat{C_{M}} || M_{safe}$ is legal for $\mathcal{M}$ with respect to $\Sigma_u \cup \mu$






        \item $\hat{C_{M}} || M_{safe} || \mathcal{M}$ is deadlock-free. \\
        Prove by contradiction. Suppose that $\hat{C_{M}} || M_{safe} || \mathcal{M}$ has a deadlock state $(c_m, m_{safe}, m) \in S_{\hat{C_{M}} || M_{safe}||\mathcal{M}}$ reachable from initial state by a trace $\pi$. \\
        Then, $\pi \in tr(\hat{C_M}) \wedge \pi \in tr(M_{cont}) \wedge \pi \in tr(\mathcal{M})$. \\
        Also, as $tr(\mathcal{M}) \subseteq tr(\mathcal{M'})$, $\pi \in tr(\mathcal{\mathcal{M'}})$ \\
        Then, there is an state $(c_m, m') \in S_{\hat{C_M} || \mathcal{M'}}$ which has at least an outgoing enabled event $l$ since $\hat{C_M}$ is solution for $\mathcal{E}$, $\hat{C_M} || \mathcal{M'}$ and it is deadlock-free. \\ 
        We  prove that $l$ is also enabled in $(c_m, m_{safe}, m) \in S_{\hat{C_{M}} || M_{safe}||\mathcal{M}}$ \\
        Suppose it is not. Then it is disabled by $m_{safe}$ or $m$ (because $c_m$ has enabled $l$ event).
        \begin{itemize}
            \item Suppose $m_{safe}$ has not enabled $l$. This is impossible because (by construction) $M_{safe}$ has \textit{safe} states and by Lemma \ref{lemma:traces-inclusion} we conclude that an event enabled in $c_m$ has to be enabled in $m_{safe}$.
            \item Suppose $m$ has not enabled $l$. This is impossible because $tr(\mathcal{M}) \subseteq tr(\mathcal{M'})$
        \end{itemize}
        Then, exists $l \in \Sigma_{\hat{C_M} || \mathcal{M'}}$ such that $l$ is enabled in $(c_m, m_{safe}, m) \in S_{\hat{C_{M}} || M_{1}^{safe}||\mathcal{M}}$. \\
        Then, $\hat{C_{M}} || M_{safe} || \mathcal{M}$ is deadlock-free.
    \end{enumerate}
    
    Then, $\hat{C_{M}} || M_{safe}$ is a solution for $\mathcal{E} = \langle \mathcal{M}, \Sigma_c, \square \Diamond \bigwedge_{l \in \mu} \neg l \implies \varphi \rangle$.

    \item $C_M || C = S$ \\
    As we assume that $C_M = \hat{C_M} || M_{safe}$ and we know that $S = \hat{C_M} || M_{safe} || C$, it trivially holds that $C_M || C = S$ using those definitions. 
\end{enumerate}

$\hfill \square$

\end{document}